\documentclass[journal]{IEEEtran}
\usepackage{graphicx, epstopdf}
\usepackage{epsfig,cite}
\usepackage{color}
\usepackage{amssymb,amsmath,mathtools, mathrsfs}
\usepackage{enumitem}
\usepackage{graphicx}
\usepackage{epstopdf}
\definecolor{blue}{rgb}{0,0,1}
\definecolor{darkgreen}{rgb}{0,.5,0}
\definecolor{darkred}{rgb}{.5,0,0}

\newtheorem{theorem}{Theorem}
\newtheorem{proposition}{Proposition}
\newtheorem{lemma}{Lemma}
\newtheorem{rem}{Remark}

\newtheorem{definition}{Definition}

\newtheorem{proper}{Property}

\def\levy{L\'evy }

\newcommand{\sysA}{\text{A}}
\newcommand{\sysB}{\text{B}}
\newcommand{\sysC}{\text{C}}

\newcommand{\realSet}{\mathcal{R}}

\def\LevyDist{{\mathscr{L}}}
\def\Pr{{\mathrm{Pr}}}

\DeclareMathOperator\erfc{erfc}
\DeclareMathOperator\gsnr{G-SNR}
\newcommand{\thr}{\mathsf{th}}
\def\sgn{\mathop{\rm sgn}\nolimits} 
\def\StabDist{{\mathscr{S}}}

\IEEEoverridecommandlockouts

\newcommand{\revised}[1]{{\leavevmode\color{black}{#1}}}
\newcommand{\nariman}[1]{{\leavevmode\color{black}{#1}}}

\begin{document}
\bstctlcite{ICC09_Ref2:BSTcontrol}

\title{Communication System Design and Analysis\\ for Asynchronous Molecular Timing Channels}

\author{ \IEEEauthorblockN{Nariman~Farsad,~\IEEEmembership{Member,~IEEE,}
		Yonathan~Murin,~\IEEEmembership{Member,~IEEE,} 
		Weisi~Guo,~\IEEEmembership{Senior Member,~IEEE,} 
	    Chan-Byoung~Chae,~\IEEEmembership{Senior Member,~IEEE,} 
		Andrew~W.~Eckford,~\IEEEmembership{Senior Member,~IEEE,} and
		\\ Andrea Goldsmith,~\IEEEmembership{Fellow,~IEEE}}
	
		\thanks{This work was been presented in part at the 2015 IEEE Global Communications Conference (GLOBECOM'2015) \cite{far15GLOBCOM}, and at the 2016 IEEE Global Communications Conference (GLOBECOM'2016) \cite{far16GLOBECOM}.}
	
	\thanks{N. Farsad, Y. Murin, and A. Goldsmith are with the Department of Electrical Engineering, Stanford University, Stanford, CA, USA, W.~Guo is with the School of Engineering, University of Warwick, Coventry, UK.  C.-B.~Chae is with the School of Integrated Technology, Yonsei University, Korea. A. Eckford is with the Department of Electrical Engineering and Computer Science, York University, Toronto, Canada.}
	\thanks{This research was supported in part by the NSF Center for Science of Information (CSoI) under grant CCF-0939370, the NSERC Postdoctoral Fellowship fund PDF-471342-2015 and by the Basic Science Research Program (2017R1A1A1A05001439) funded by the MSIP, Korea, through the NRF of Korea. }
}

%
\maketitle

\begin{abstract}
Two new asynchronous modulation techniques for molecular timing (MT) channels are proposed. One based on modulating information on the time between two consecutive releases of {\em indistinguishable} information particles, and one based on using {\em distinguishable} particles. For comparison, we consider the synchronized modulation scheme where information is encoded in the time of release and decoded from the time of arrival of particles. We show that all three modulation techniques result in a system that can be modeled as an additive noise channel, and we derive the expression for the probability density function of the noise. Next, we focus on binary communication and derive the associated optimal detection rules for each modulation. Since the noise associated with these modulations has an infinite variance, geometric power is used as a measure for the noise power, and we derive an expression for the geometric SNR (G-SNR) for each modulation scheme. Numerical evaluations indicate that for these systems the bit error rate (BER) is constant at a given G-SNR, similar to the relation between BER and SNR in additive Gaussian noise channels. We also demonstrate that the asynchronous modulation based on two {\em distinguishable} particles can achieve a BER performance close to the synchronized modulation scheme.
\end{abstract}

\begin{IEEEkeywords}
Molecular communication, channel models, noise models, L\'evy distribution, stable distributions, bit error rate, and molecular timing channel.
\end{IEEEkeywords}

\section{Introduction}

\IEEEPARstart{M}{olecular} communication is a biologically inspired form of communication, where chemical signals are used to transfer information \cite{far16ST,mole_book,weimag}. It is possible to modulate information on the particles using different techniques such as concentration \cite{kur12}, type, ratio \cite{kim13}, number \cite{far12NanoBio}, time of release \cite{eck09}, or a combination of these techniques \cite{far16ST}. Moreover, information particles can be transported from the transmitter to the receiver using diffusion \cite{mah10}, active transport \cite{far11Bionet},  bacteria \cite{lio12}, \revised{and/or flow} (or advection) \cite{bic13}. Between all these techniques, diffusion and flow-based propagation are the easiest to implement, and a few experimental platforms have been built to demonstrate molecular communication based on these transport mechanisms~\cite{far13,far14INFOCOM,koo16}.

In this work we focus on modulation techniques for molecular communication and their corresponding system models. Most prior work on modulation techniques rely on the concentration or the type of the released particles. For example in \cite{ata12}, the order of release of consecutive distinguishable particles is proposed for encoding information. In this work, we consider molecular timing (MT) channels where timing-based modulation is employed. Only a few works have considered this type of modulation: In \cite{eck09} the time of release of the particles is used for encoding information, while in \cite{kri13} the information is encoded in the time interval between two pulse releases of information particles \revised{in mirofluidic channels}.

The work \cite{sri12} showed that in the case of timing-based modulation, where information is encoded on the release timing of particles, and the transport mechanism is diffusion assisted by constant laminar flow, the channel can be represented as an additive noise channel. In this case the noise term follows the inverse Gaussian (IG) distribution. 
Capacity bounds for the additive IG noise channel, in bits per channel use, under an average delay constraint, were derived in \cite{sri12, li14}. 
In \cite{far15GLOBCOM}, we have shown that in the case of timing-based modulation, where information is encoded on the release timing of particles, and pure diffusive transport (i.e., diffusion without any flow) is employed, the channel can be represented as an additive noise channel where the noise follows the L\'evy distribution. 
The capacity of this channel was studied in \cite{far16ISIT, rose16Part1, rose16Part2, far17TIT}, and it was shown that this capacity can increase poly-logarithmically with respect to the number of simultaneously released particles. A sequence detector for this modulation scheme was presented in \cite{mur17NanoComNet}.

\begin{figure}
	\begin{center}
		\includegraphics[width=\columnwidth]{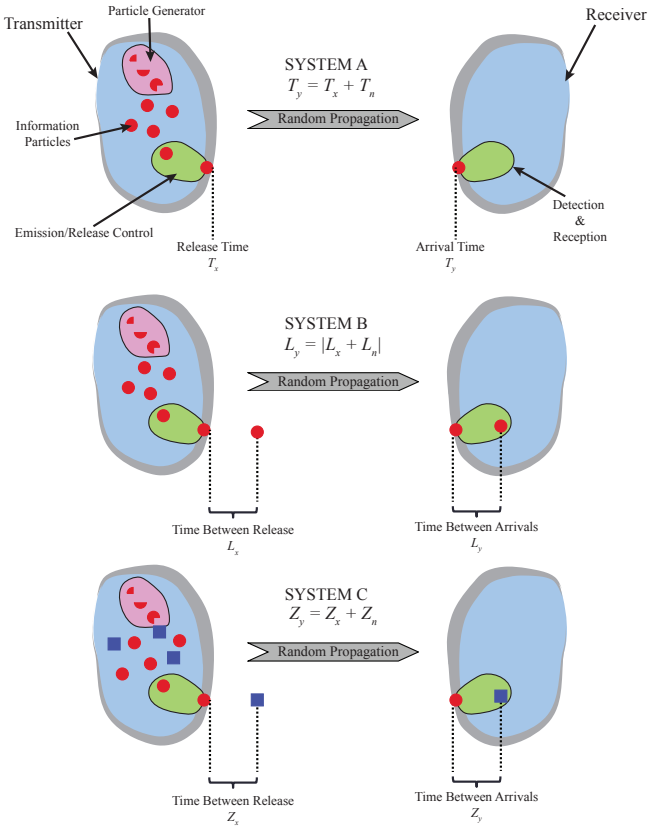}
	\end{center}
	\vspace{-0.3cm}
	\caption{\label{fig:chanModels} Summary of the System models corresponding to each modulation scheme. System A: a synchronized MT, system B: an asynchronous MT with indistinguishable information particles, and system C: an asynchronous MT with distinguishable information particles }
	\vspace{-0.3cm}
\end{figure}

\revised{In this work, we propose two asynchronous timing-based modulation techniques and compare them with the synchronized timing modulation considered in prior work \cite{mur17NanoComNet,mur16Globecom,mur17}.} 
These systems can be represented by an additive noise channel, and for diffusion-based MT (DBMT) systems \revised{in a 1-dimensional environment} the noise falls in the stable distribution family~\cite{nol15}. \revised{The models considered in this paper can be used to represent molecular channels for communication on bio-chips. In bio-chips, components within a chip, such as a storage unit, a molecular processing unit, sorting unit, etc. are connected by narrow microfluidic links. Since these links are very narrow, they are well approximated as a 1-dimensional environment. Moreover, the transmitter and the receiver in these applications can be designed with great precision \cite{lee09,hor15}. 

The three systems considered in this paper are as follows. First, we consider a synchronized MT system, where information is encoded in the release timing of information particles (system A); second, an asynchronous MT system is proposed where information is encoded in the time between two consecutive releases of {\em indistinguishable} information particles (system B); and finally, another asynchronous MT system is considered where information is encoded in the time between two consecutive releases of {\em distinguishable} information particles (system C).} Fig.~\ref{fig:chanModels} depicts all three systems. 
One of the main motivations for proposing these new modulations is the challenge of synchronization.  In particular, for some applications involving micro and nano-scale devices, it may be difficult to synchronize the transmitter and the receiver due to their small size and limited power. In this case, the modulation scheme in system A, which has been used in previous works, may be too difficult to implement in practice. The newly presented modulation schemes in systems B and C, however, do not require synchronization between the transmitter and the receiver. These modulations are analogous to differential phase-shift keying (PSK) in that the asynchronous MT modulations do not require an absolute time reference, while the differential PSK does not require an absolute phase reference.

It must be noted that stable distributed noise arises in system models for a number of different applications. Therefore, the results of this paper could also be applicable in those areas. Specifically, in \cite{he14}, alpha-stable distributed noise was used to model room acoustics. In radio communications, symmetric alpha-stable distributions were used to model impulsive non-Gaussian noise such as those that exists in ultra-wide bandwidth systems~\cite{nir09,fan12}. Capacity bounds for a special class of alpha-stable additive noise channels were provided in~\cite{wan11,fah12}. \revised{Although in this work we focus on additive stable distributed noise channels in the context of molecular timing channels, the analysis and the results are applicable to the general detection problem in additive stable distributed noise channels.}

There are only three classes of stable distributions with closed-form probability density functions (PDF) in terms of elementary functions: Gaussian, Cauchy, and L\'evy. In this work, we derive closed-form expressions for the PDFs of the noise terms in systems B and C in terms of the complex error function and Voigt functions \cite{voigt92, abr11}, which are used in other fields of science such as physics. Thus we develop new closed-form PDF results for a subclass of stable noise distributions in terms of the Voigt functions, which can be efficiently calculated numerically \cite{zag11, abr15}, and can be approximated using elementary functions in some special cases \cite{sch16}. 

To compare the performance of the three proposed modulation schemes, we consider a binary communication system and derive the optimal detection rule for each modulation technique. Since the system noise in all three cases is heavy-tailed with infinite variance, the standard definition of signal power, used in electromagnetic communication, is not suitable. Instead, we derive the expressions for the geometric power \cite{gon06} of a large class of stable distributions, and use it to represent the noise power. Furthermore, instead of using the well known signal-to-noise ratio (SNR) metric, we use the geometric SNR (G-SNR)~\cite{gon06} metric, which is given by the geometric power of the signal divided by the geometric power of the noise with some normalization constants. Based on numerical evaluations we observe that for the modulations considered, the bit error rate (BER) is constant for a given G-SNR regardless of the geometric signal power and the geometric noise power. 

Based on the above derivations, we next use numerical evaluations to compare the BER of all three systems. We show that system B with indistinguishable particles exhibits the highest BER, while system A achieves the lowest BER. This indicates that time-synchronized transmission over MT channels, i.e. Modulation A, works better than the other two modulations considered.
We further show that by adjusting the diffusion coefficients of the information particles in system C, which is an asynchronous transmission, the BER can approach the BER of system A, where full synchronization is assumed. 
However, this comes at the cost of added system complexity where both the transmitter and receiver must be capable of transmitting and detecting two distinguishable particles.  

The rest of this paper is organized as follows. In Section \ref{sec:model} we present the three timing-based modulation techniques, and derive an additive noise system model for each of them. In Section \ref{sec:noiseModel} we focus on the diffusion-based propagation and derive the PDF for the additive noise term for each system. In Section \ref{sec:PBE}, binary communication is studied, and the optimal detectors are derived. The geometric power of the noise and the G-SNR of each system are derived in Section \ref{sec:SNR}.  
Numerical BER evaluations of the proposed modulation techniques are presented in Section \ref{sec:numEval}, and concluding remarks are provided in Section \ref{sec:conc}.   

\textbf{\textit{Notation:}} We denote the set of real numbers by $\realSet$, and the set of positive real numbers by $\realSet^{+}$. Other than these sets, we denote sets with calligraphic letters, e.g., $\mathcal{T}$. We denote random variables (RV)s with upper case letters, e.g., $X$ and $Y$, and their realizations with the corresponding lower case letters, e.g., $x$ and $y$. We use $f_{Y}(y)$ to denote the PDF of a continuous RV $Y$ on $\realSet$, $f_{Y|X}(y|x)$ to denote the conditional PDF of $Y$ given $X$, and  $F_{Y}(y)$ and $F_{Y|X}(y|x)$ to denote the corresponding cumulative distribution functions (CDF).  We use $\varphi_{X} (\revised{\omega})$ to denote the characteristic function of the RV $X$ and we use the notation $X \overset{d}{=} Y$ to denote the equality in distribution, i.e., $X$ has the same PDF as $Y$. 
We use $|\cdot|$ to denote the absolute value, $j \triangleq \sqrt{-1}$ to denote the imaginary number, and $\Re\{ z \}$ to denote the real part of the complex number $z$.  
Finally, $\erfc\left( \cdot \right)$ is used to denote the complementary error function given by $\erfc(x) = \frac{2}{\sqrt{\pi}} \int_{x}^{\infty}{e^{-u^2} du}$.

\section{System Models}
\label{sec:model}
In this section we present three different timing-based modulation techniques, which results in three different MT system models. Note that there are no unified channel models for all possible modulation schemes in molecular communication, and typically each modulation yields a different channel model. To develop our model, we make the following assumptions about the system:
\begin{enumerate}[label = {\bf A{\arabic*}})]
	\item \label{assmp:perfectTxRx}
	The transmitter perfectly controls the release time of each information particle, and the receiver perfectly measures the arrival times of the information particles. 
	Furthermore, the transmitter and the receiver are perfectly synchronized in time, when synchronization is required by the modulation scheme.
	
	\item \label{assmp:Arrival}
	Any information particle that arrives at the receiver is absorbed and hence is removed from the propagation medium.
	
	\item \label{assmp:indep}
	All information particles propagate independently of each other, and their trajectories are random according to an independent and identically distributed (i.i.d.)  random process. This is a fair assumption for many different propagation schemes in molecular communication such as diffusion in dilute solutions, i.e., when the number of particles released is much smaller than the number of molecules of the solutions.
	
	\item \label{assmp:noISI}
	There is no inter-symbol interference (ISI) between consecutive channel uses.\footnote{In \cite{mur17NanoComNet} we study communication over DBMT channels in the presence of ISI.} In practice this assumption can be satisfied if the time between consecutive channel uses is large enough or if chemical reactions are used to dissipate the particles \cite{noe14}.
    
    \item 
    \revised{Initially, there are no information particles in the environment and the particles that arrive at the receiver are the ones released by the transmitter.} 
\end{enumerate} 
\noindent \revised{Note that these assumptions are typical in the study of molecular communication systems \cite{ein2011,ein11ITW,ata13,nak12, cha12,eck12,pie13, li14, rose14}. 

Moreover, as our system uses molecular timing channels through microfluidic links on bio-chips, the assumptions can be satisfied as follows. In bio-chips, it is possible to release and detect information particles with great precision \cite{lee09,hor15} (satisfies A1). Many detectors will remove the information particles as part of the detection process, e.g., by binding to a receptor, diffusion into a sensing layer, or chemical reactions (satisfies A2). The number of information particles released as part of the modulation techniques considered in this work are one or two (satisfies A3). The applications considered in this work are biological studies, where only a very few bits of information need to be transmitted followed by large periods of silence (satisfies A4). Finally, the microfluidic links can be carefully monitored and designed to exclude interfering molecules (satisfies A5).} 

The first MT system that we consider is the one proposed in \cite{eck09,sri12}, where the information is encoded in the release timing of a single information particle. Let $T_{x} \in \mathcal{T}\subseteq\realSet^+$ be the release time of the information particle at the transmitter. In this scheme, the information is modulated onto the release time itself. The released particle is then transported from the transmitter to the receiver, where the transport process is random. Let $T_{y}$ be the time of arrival at the receiver. Then we have
\begin{align}
\label{eq:timingChA}
	T_{y} = T_{x} + T_n,
\end{align}
where $T_n$ is the random propagation delay of the information particle, which is the {\em noise term} in this channel. One of the main challenges of this modulation scheme is the need for synchronization between the transmitter and the receiver. In this work, whenever system A is used we assume that the transmitter and the receiver are perfectly synchronized.

To overcome this synchronization challenge, we propose two new modulation schemes in which information is modulated on the time duration {\em between} two consecutive releases of information particles. The receiver decodes the information from the time between the arrivals of two molecules. Note that in this case synchronization between the transmitter and the receiver is not required.
Two cases are possible: either the two released information particles are {\em indistinguishable} at the receiver, or the two released information particles are {\em distinguishable} at the receiver. 

We first consider the case where both information particles are {\em indistinguishable}. Without loss of generality, let $T_{x_1}$ be the release timing of the first information particle and let $T_{x_2}$ be the release timing for the second information particle, with $T_{x_2}>T_{x_1}$. Thus, the information is encoded in $L_x = T_{x_2}-T_{x_1}$. Using \eqref{eq:timingChA}, the system model for this modulation scheme is given by:
\begin{align}
\label{eq:timingChB}
	|T_{y_2} - T_{y_1}|&= |T_{x_2}-T_{x_1} + T_{n_2} - T_{n_1}|,  \nonumber \\
	L_y &= | L_x+L_n|, 
\end{align}
where $L_n = T_{n_2} - T_{n_1}$ is the random noise term in this system, and $T_{n_2}$ and $T_{n_1}$ are the random propagation delays for the first and the second particles as in (\ref{eq:timingChA}). Note that the absolute value in the system formulation is due to the fact that both information particles are indistinguishable, and therefore the receiver can observe only the absolute difference of arrival times.

The last modulation scheme uses the time between releases of two {\em distinguishable} information particles (i.e., two different particle types) to encode information. Let $T_{x}^a$ be the release timing of the type-$a$ information particle and let $T_{x}^b$ be the release timing of the type-$b$ information particle. We assume that the information is encoded in $Z_x = T_{x}^b-T_{x}^a$. Unlike (\ref{eq:timingChB}) where $L_x$ is always positive, $Z_x$ can be positive or negative depending on the order that the type-$a$ and type-$b$ information particles are released. Using (\ref{eq:timingChA}), the system model for this scheme is given by:
\begin{align}
\label{eq:timingChC}
	T_{y}^b - T_{y}^a &= T_{x}^b-T_{x}^a + T_{n}^b - T_{n}^a,  \nonumber \\
	Z_y &= Z_x+ Z_n, 
\end{align}
where $Z_n = T_{n}^b - T_{n}^a$ is the random additive noise term in this system, and $T_{n}^b$ and $T_{n}^a$ are the random propagation delays for the type-$a$ and type-$b$ particles as in (\ref{eq:timingChA}). Again, no synchronization is required between the transmitter and receiver. Fig.~\ref{fig:chanModels} summarizes all three modulation techniques.

Note that the proposed modulation schemes and their corresponding system models could be applied to any type of propagation model through the medium as long as Assumptions \ref{assmp:perfectTxRx}-\ref{assmp:noISI} are not violated. In the next section, we derive the distribution of the noise terms for the proposed MT systems, when diffusion propagation is used for particle transport.

\section{Noise Models for DBMT Systems}
\label{sec:noiseModel}
In the rest of this work we focus on DBMT systems where diffusion is used for particle transport. In particular, in this section we derive the PDF of the noise terms $T_n$, $L_n$, and $Z_n$ for DBMT systems in \eqref{eq:timingChA}-\eqref{eq:timingChC}, and discuss some of the properties of these RVs.

\subsection{System A}

To specify the random additive noise term $T_n$ in system A, we define a L\'evy-distributed RV as follows. 
\begin{definition}[\levy Distribution] \label{def:levyRV}
	Let the RV $X$ be L\'evy-distributed with location parameter $\mu$ and scale parameter $c$ \cite{nol15}. Then its PDF is given by:
	\begin{align}
	\label{eqn:LevyPDF_0}
	f_X(x;\mu,c)=
	\begin{cases}
	\sqrt{\frac{c}{2 \pi (x-\mu)^3}}\exp \left( -\frac{c}{2(x-\mu)} \right), & x>\mu \\
	0, & x\leq \mu
	\end{cases},
	\end{align} 
	its characteristic function is given by:
	\begin{align}
	\varphi (\revised{\omega};\mu,c) = \exp \left( j \mu \revised{\omega} -\sqrt{-2 j c \revised{\omega} } \right),
	\end{align}
	and its CDF is given by:
	\begin{align}
	\label{eqn:LevyCDF}
	F_X(x;\mu,c) = \begin{cases} \erfc\left(\sqrt{\frac{c}{2(x-\mu)}}\right), & x>\mu \\ 0, & x\leq\mu \end{cases}.		
	\end{align}
\end{definition}
Throughout the paper, we use the notation $X \sim \LevyDist(\mu,c)$ to indicate a \levy RV with parameters $\mu$ and $c$.

Assumption \ref{assmp:Arrival} implies that the distribution of $T_n$ is the distribution of the first hitting time (first arrival at the receiver) of a particle transported via diffusion without flow. 
In previous works, it was shown that the first hitting time for a diffusion channel with constant drift (i.e., flow) in 1-dimensional space follows the inverse Gaussian distribution \cite{sri12}. In this work, we consider the pure diffusion channel with no flow. Let $d$ denote the distance between the transmitter and the receiver, and $D$ denote the diffusion coefficient of the information particles in the propagation medium. Following along the lines of the derivations in \cite[Sec. II]{sri12}, and using \cite[Sec. 2.6.A]{karatzas-shreve}, it can be shown that for 1-dimensional pure diffusion, the propagation time of each of the information particles follows a \levy distribution, and therefore the noise in system A is distributed as $T_n \sim \LevyDist(0,c_{\sysA})$ with $c_{\sysA} = \frac{d^2}{2D}$. Similarly, the conditional PDF $P(T_{y}|T_{x}) \sim \LevyDist(T_{x},c_{\sysA})$. 
The PDF and CDF of the standardized (i.e., with $\mu=0$ and $c_{\sysA} =1$) \levy noise are depicted in Figs. \ref{fig:stablePDF} and \ref{fig:stableCDF}, respectively. 

\begin{rem}
	In \cite{yilmaz20143dChannelCF} it is shown that for an infinite,  three-dimensional homogeneous medium without flow, and a spherically absorbing receiver, the first arrival time follows a scaled \levy distribution. Therefore, the results presented in this paper can be extended to 3-D space by simply introducing a scalar multiple in the noise distribution. 
\end{rem}

\subsection{System B}

To find the noise distribution of the system in (\ref{eq:timingChB}), we first discuss the class of probability distributions known as {\em stable distributions} \cite{zol86-book,nol15}. Note that the L\'evy distribution belongs to this class.
\begin{definition}[Stable Distributions]
	\revised{A} RV $X$ has a stable distribution if for two independent copies $X_1$ and $X_2$, and positive constants $a_1, a_2, a_3 \in \realSet^{+}$ and $a_4 \in \realSet$, the following holds:
\begin{align*}
	a_1X_1 + a_2X_2 \overset{d}{=} a_3X+a_4.
\end{align*}

\end{definition}

Stable distributions can also be defined via their characteristic function.
\begin{definition}[Characteristic Function of a Stable Distribution]
	Let $-\infty < \mu < \infty, c\ge 0, 0 < \alpha \le 2$, and $-1 \le \beta \le 1$. Further define: 
	\begin{align*}
		\Phi(\revised{\omega},\alpha) \triangleq \begin{cases} 
												\tan \left( \frac{\pi \alpha}{2}\right), & \alpha \ne 1 \\ 
												-\frac{2}{\pi} \log (|\revised{\omega}|), & \alpha = 1
											\end{cases}.
	\end{align*}
	
	\noindent Then, the characteristic function of a stable RV $X$, with location parameter $\mu$, scale parameter $c$, characteristic exponent $\alpha$, and skewness parameter $\beta$, is given by:
	\begin{align}
	\varphi (\revised{\omega};\mu,c,\alpha,\beta) = \exp \left[ j \mu \revised{\omega} -| c \revised{\omega} |^\alpha (1-j\beta \sgn(\revised{\omega}) \Phi(\revised{\omega}, \alpha) ) \right].
	\label{eq:stableCharFunc}
\end{align}
\end{definition}

In the following, we use the notation $\mathscr{S}(\mu, c, \alpha, \beta)$ to represent a stable distribution with the parameters $\mu, c, \alpha$, and $\beta$.
Only the PDFs of three classes of stable distributions are known to have closed-form expressions in terms of elementary functions: the Gaussian distribution with $\alpha=2$ (the value of $\beta$ does not matter in this case and can be assumed to be zero), the L\'evy distribution with $\alpha = \tfrac{1}{2}$ and $\beta=1$, and the Cauchy distribution with $\alpha=1$ and $\beta=0$. Generally, the parameters $\alpha$ and $\beta$ define a subclass within the stable distribution family. Next, we introduce some important properties of stable distributions \cite{nol15}.

\begin{proper}
\label{prop:LinearStable}	
	Let $X \sim \StabDist(\mu,c,\alpha,\beta)$, and define $Y = \frac{X-\mu}{c}$. Then $f(x)dx =f(y)dy$, and $Y$ is called the standard form of $X$.	
\end{proper}
\begin{proper}
	\label{prop:standStable}	
	Let $\tilde{X} \sim \StabDist(0,1,\alpha,\beta)$ be the standard form of a stable RV with parameters $\alpha$ and $\beta$. Then the PDF and the CDF of any RV $X \sim \StabDist(\mu,c,\alpha,\beta)$ can be calculated as
	\begin{align}
	f_X(x) &= \frac{f_{\tilde{X}}\big( \tfrac{x-\mu}{c} \big)}{c}, \label{eq:standPDFConv}\\
	F_X(x) &=  F_{\tilde{X}}( \tfrac{x-\mu}{c} ). \label{eq:standCDFConv}
	\end{align}
\end{proper}
Using this property, the standard PDF and CDF of a stable RV can be used to calculate probabilities involving non-standard stable RVs just like the way the standard Gaussian PDF and CDF are used to calculate probabilities involving non-standard Gaussian RVs.
\begin{proper}
	\label{prop:sym}
	The PDFs of stable RVs with $\beta=0$ are symmetric around $\mu$.	
\end{proper}
\begin{proper}
	\label{prop:tailProbability}
	If $X$ is a standardized (i.e., with $\mu=0$ and $c =1$) stable RV with parameters $0<\alpha<2$ and $\beta$, then as $x\rightarrow\infty$,
	\begin{align}
	P(X>x;\alpha,\beta) \approx \frac{1+\beta}{\pi x^\alpha} \Gamma(\alpha)\sin\bigg(\frac{\alpha \pi}{2}\bigg). 
	\end{align} 
\end{proper}
\begin{rem}
	Using this property it can be shown that for a stable distributed RV $X$ with parameter $\alpha$, the moments of order greater than $\alpha$ (i.e., $\mathbb{E}[|X|^\alpha]$) are infinite. Therefore, all stable distributions with $\alpha<2$ have infinite variances, and all stable distributions with $\alpha <1$ have infinite mean values. 
\end{rem}

With these definitions we now model the noise term $L_n$ in (\ref{eq:timingChB}). 
\begin{theorem}
Let $c_{\sysB} = \frac{2d^2}{D} $, where $d$ is the distance between the transmitter and the receiver and $D$ is the diffusion coefficient of the information particles. Then, the characteristic function of the noise term $L_n$ is given by:
\begin{align*}
\varphi \left( \revised{\omega};c_{\sysB} \right) =  \exp \left[ -\sqrt{c_{\sysB}| \revised{\omega} |} \right],
\end{align*}
which implies that $L_n \sim \StabDist(0,c_{\sysB},\frac{1}{2},0)$.
\end{theorem}
\begin{IEEEproof}
We know that $L_n = T_{n_2}+( - T_{n_1})$ with $T_{n_2}, T_{n_1} \sim \StabDist(0,c_{\sysA},\frac{1}{2},1)$, where $c_{\sysA} = \frac{d^2}{2D}$. Since $T_{n_1}$ and $T_{n_2}$ are independent, the characteristic function for $L_n$ is given by
\begin{align}
\varphi_{L_n} (\revised{\omega}) &=  \varphi_{T_{n_2}} (\revised{\omega}) \varphi_{T_{n_1}} (-\revised{\omega}) \\
						&= \exp \left[ -\sqrt{| c_{\sysA} \revised{\omega} |} (1- j \sgn(\revised{\omega})) \right]  \times \nonumber \\ 
						& ~~~~~~~~~~~~~~~~~~~~\exp \left[ -\sqrt{| c_{\sysA} \revised{\omega} |} (1+ j \sgn(\revised{\omega})) \right] \\
						&= \exp \left[ -\sqrt{|4c_{\sysA} \revised{\omega} |}  \right].
\end{align}

\noindent Thus, using the expression in \eqref{eq:stableCharFunc} we conclude that $L_n \sim \StabDist(0,c_{\sysB},\frac{1}{2},0)$.
\end{IEEEproof}

\begin{rem}
	If the same type of particle is used in system A and system B, and the distance between the transmitter and the receiver is the same, then the scale parameter $c$ in the noise term for system B is four times greater than the scale parameter for system A, i.e., $c_{\sysB} = 4c_{\sysA}$. 
\end{rem}

\begin{figure}
	\begin{center}
		\includegraphics[width=\columnwidth]{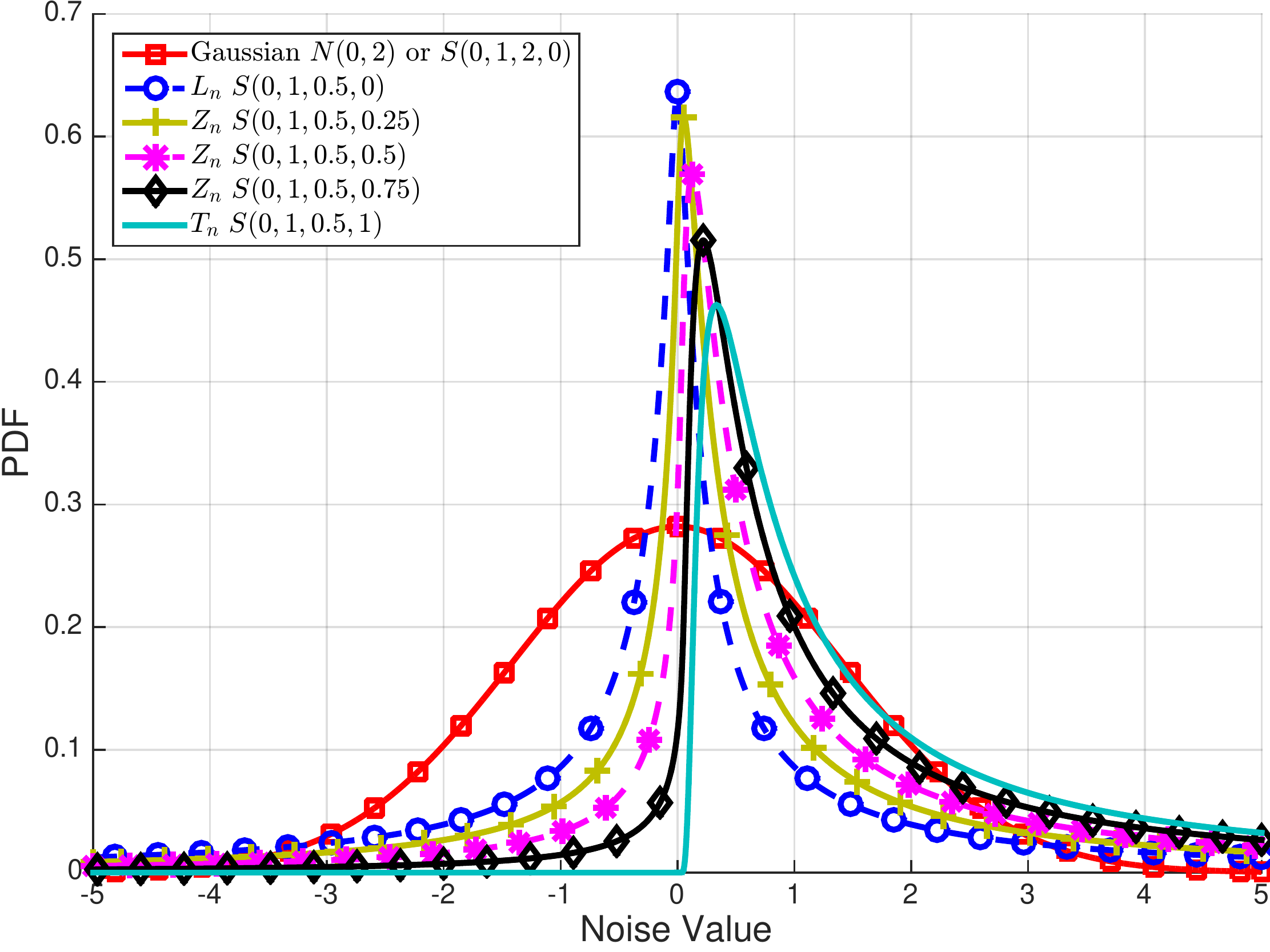}
	\end{center}
	\caption{\label{fig:stablePDF} The probability density function of different standardized noise terms.}
\end{figure}
\begin{figure}
	\begin{center}
		\includegraphics[width=\columnwidth]{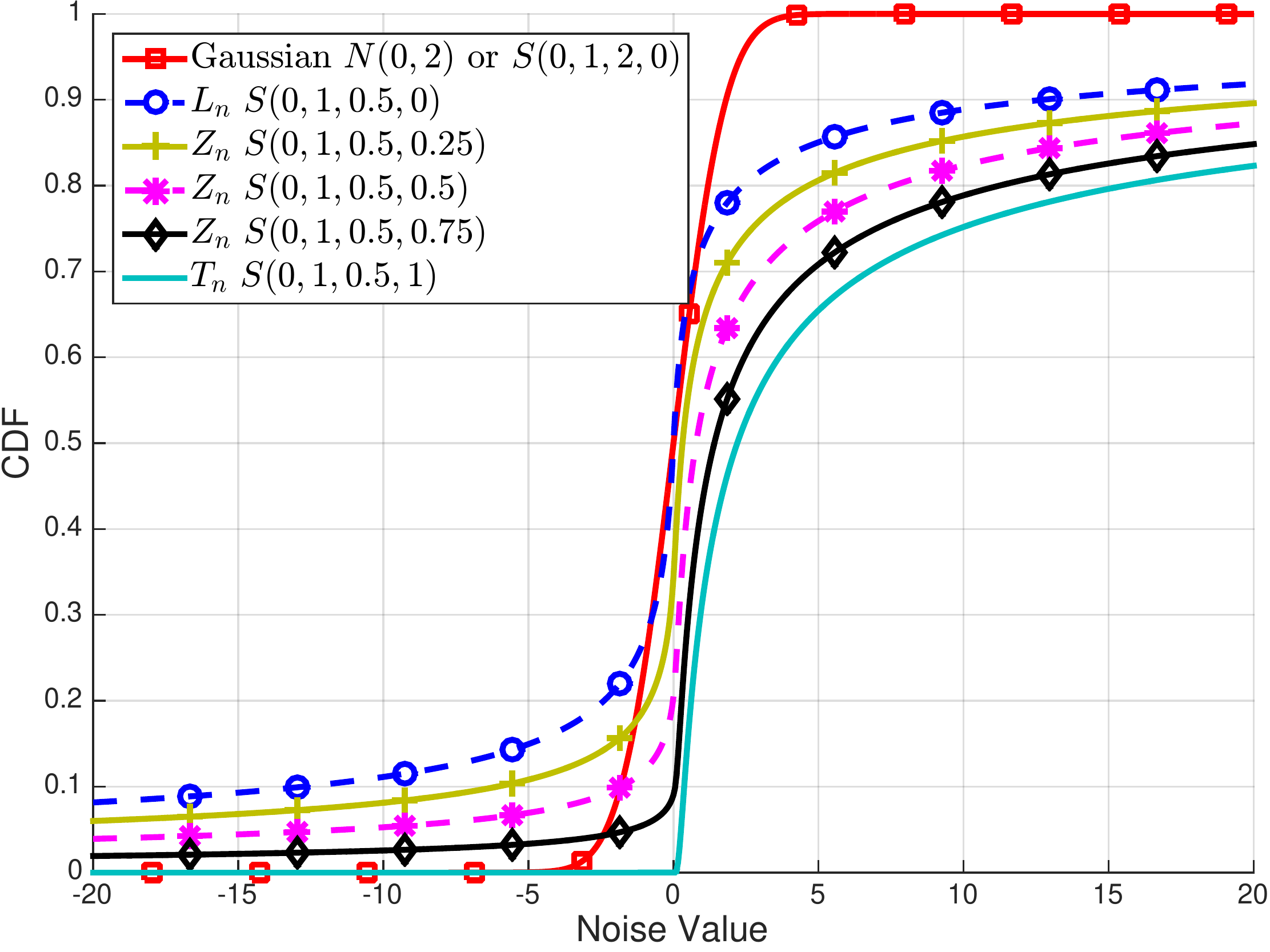}
	\end{center}
	\caption{\label{fig:stableCDF} The cumulative distribution function of different standardized noise terms.}
\end{figure}

To find an expression for the PDF of the noise term $L_n$ in (\ref{eq:timingChB}), we first define the following functions. 
 Let $K(a,b)$ and $L(a,b), a \in \realSet, b \in \realSet^{+}$, be the complex and imaginary Voigt functions \cite{arm67}, given by :
\begin{align}
\label{eq:ReVoigt}
K(a,b) \triangleq \frac{1}{\sqrt{\pi}} \int_0^\infty \exp(-t^2/4)\exp(-bt)\cos(at) dt, 
\end{align}
and
\begin{align}
\label{eq:ImVoigt}
L(a,b) \triangleq \frac{1}{\sqrt{\pi}} \int_0^\infty \exp(-t^2/4)\exp(-bt)\sin(at) dt.
\end{align}
The Voigt functions, which are widely used in the fields of physics, astronomy, and chemistry, can be computed efficiently and quickly numerically \cite{zag11, abr15}. Moreover, for some special cases (e.g., $b\gg0$), analytical approximations of these functions exist in terms of elementary functions \cite{sch16}. We further define:
\begin{align}
\label{eq:theGfunc}
G(u) &\triangleq \frac{1}{\sqrt{8\pi |u|^3}} \left[ K \left(-\tfrac{1}{\sqrt{8|u|}}, \tfrac{1}{\sqrt{8|u|}} \right) \right. \nonumber \\
 &\qquad \qquad \qquad \left. + L\left(-\tfrac{1}{\sqrt{8|u|}}, \tfrac{1}{\sqrt{8|u|}} \right)  \right]. 
\end{align}

The PDF of $L_n$ is stated in the following theorem:
\begin{theorem}
	\label{thm:PDFNoiseChanB}
	Let $L_n \sim \StabDist(0,c_{\sysB},\frac{1}{2},0)$. Then the PDF of $L_n$ is given by:
	\begin{align}
	\label{eq:noisePDFchanB}
		f_{L_n}(\ell_n) = \begin{cases} 
		\frac{1}{c_{\sysB}} G\left( \tfrac{\ell_n}{c_{\sysB}} \right), & \ell_n \neq 0 \\
		\frac{2}{c_{\sysB}\pi}, & \ell_n = 0  
		\end{cases}. 
	\end{align}
\end{theorem}

\begin{IEEEproof}
	The proof is provided in Appendix \ref{annex:ProofOfPDFchanB}.
\end{IEEEproof}

In this work, we do not provide an expression for the CDF of the noise terms $F_{L_n}(\ell_n)$, and it would be difficult to integrate \eqref{thm:PDFNoiseChanB} to obtain the CDF. However, the CDF can be calculated numerically using the methods described in \cite[Sec. 3]{nol97}. Moreover, tables of the standardized CDF could be used to calculate probabilities involving the noise term. Figs. \ref{fig:stablePDF} and \ref{fig:stableCDF} depict the PDF and CDF for the standardized noise term $L_n$ with $c_{\sysB} =1$. 

\subsection{System C}

We first note that the noise $Z_n$ given in (\ref{eq:timingChC}) is fundamentally different from the noise $L_n$ in system B since the two different types of information particles may have different diffusion coefficients. Let $D_a$ be the diffusion coefficient of information particle $a$, and $D_b$ be the diffusion coefficient for the information particle $b$. We define $c_{\sysC} \triangleq \frac{d^2(\sqrt{D_a}+\sqrt{D_b})^2}{2D_aD_b}$,  $\beta_{\sysC} \triangleq  \frac{\sqrt{D_a}-\sqrt{D_b}}{\sqrt{D_a}+\sqrt{D_b}}$. Furthemore, without loss of generality, we assume that particle $a$ is released before particle $b$. We now model the noise term $Z_n$ in (\ref{eq:timingChC}).
\begin{theorem}
The characteristic function for the noise term $Z_n$ is given by: 
\small
\begin{align*}
&\varphi \left( \revised{\omega};c_{\sysC}, \beta_{\sysC}\right) =  \exp \left[ -\sqrt{c_{\sysC}| \revised{\omega} |} \left( 1- j\beta_{\sysC}\sgn(\revised{\omega})    \right) \right],
\end{align*}
which implies that $Z_n \sim \StabDist \left( 0,c_{\sysC},\frac{1}{2},\beta_{\sysC} \right)$.
\end{theorem}

\begin{IEEEproof}
First, note that $Z_n = T_{n_b}+( - T_{n_a})$ with $T_{n_a}, T_{n_b} \sim \StabDist(0,c_i,\frac{1}{2},1)$, where $c_i = \frac{d^2}{2D_i}$ for $i \in\{ a,b \}$. Since $T_{n_a}$ and $T_{n_b}$ are independent, the characteristic function of $Z_n$ is given by:
\begin{align}
& \varphi_{Z_n} (\revised{\omega}) \nonumber \\
& \quad =  \varphi_{T_{n_b}} (\revised{\omega}) \varphi_{T_{n_a}} (-\revised{\omega}) \\
						& \quad = \exp \left[ -\sqrt{c_b}\sqrt{| \revised{\omega} |} (1- j \sgn(\revised{\omega})) \right]  \times \nonumber \\ 
						& \mspace{100mu} \exp \left[ -\sqrt{c_a}\sqrt{| \revised{\omega} |} (1+ j \sgn(\revised{\omega})) \right] \\
						& \quad = \exp \left[ - \mspace{-3mu} \sqrt{|\revised{\omega}|}(\sqrt{c_b} \mspace{-3mu} + \mspace{-3mu} \sqrt{c_a} \mspace{-3mu} - \mspace{-3mu} j\sgn(\revised{\omega})(\sqrt{c_a} \mspace{-3mu} - \mspace{-3mu} \sqrt{c_b}))  \right] \\
						& \quad = \exp \left[ -(\sqrt{c_b}+\sqrt{c_a})\sqrt{|\revised{\omega}|} \times \right. \nonumber \\ 
						& ~~~~~~~~~~~~~~~\left. \left( 1-j\sgn(\revised{\omega})\frac{\sqrt{c_a}-\sqrt{c_b}}{\sqrt{c_b}+\sqrt{c_a}}\right)  \right] \\
						& \quad = \exp \left[ -\frac{d(\sqrt{D_a}+\sqrt{D_b})}{\sqrt{2D_aD_b}}\sqrt{| \revised{\omega} |} \times \right. \nonumber \\ 
		&~~~~~~~~~~~~~~~~~~~~\left. \left( 1- j \frac{\sqrt{D_a}-\sqrt{D_b}}{\sqrt{D_a}+\sqrt{D_b}}\sgn(\revised{\omega}) \right)\right].
\end{align}

\noindent Thus, using the expression in \eqref{eq:stableCharFunc} we conclude that $Z_n \sim \StabDist \left( 0,c_{\sysC},\frac{1}{2},\beta_{\sysC} \right)$.
\end{IEEEproof} 

\begin{rem}
	When the diffusion coefficients of the two particles are approximately the same, i.e., $D_a \approx D_b$, the distribution of $Z_n$ approaches the distribution $L_n$ (i.e. $\beta_{\sysC} \approx 0$). On the other hand, when $D_a \ll D_b$ or $D_a \gg  D_b$, then $\beta_{\sysC} \approx \pm1$ which implies that $Z_n$ is L\'evy distributed. 
Therefore, when one information particle has a much higher diffusion coefficient than the other, system C can be reduced to system A with the added benefit that no synchronization is required between the transmitter and the receiver. 
However, this comes at a cost of: 1) Using two particles instead of one; and 2) The resulting system A has a scaling parameter that corresponds to the smaller diffusion coefficient.
\end{rem}

To derive the PDF of $Z_n$ we first define the following two functions:
\begin{align}
G_+(u,\beta) &\triangleq \frac{1}{\sqrt{8\pi |u|^3}} \left[ (1+\beta)K \left(-\tfrac{1+\beta}{\sqrt{8|u|}}, \tfrac{1-\beta}{\sqrt{8|u|}} \right) \right. \nonumber \\
&\qquad \left. + (1-\beta)L\left(-\tfrac{1+\beta}{\sqrt{8|u|}}, \tfrac{1-\beta}{\sqrt{8|u|}} \right)  \right], \label{eq:theGpfunc}\\ 
G_-(u,\beta) &\triangleq \frac{1}{\sqrt{8\pi |u|^3}} \left[ (1-\beta)K \left(\tfrac{1-\beta}{\sqrt{8|u|}},\tfrac{1+\beta}{\sqrt{8|u|}} \right) \right. \nonumber \\
&\qquad \left. + (1+\beta)L\left(\tfrac{1-\beta}{\sqrt{8|u|}}, \tfrac{1+\beta}{\sqrt{8|u|}} \right)  \right], \label{eq:theGnfunc}
\end{align}
where $K(a,b)$ and $L(a,b)$ are the real and imaginary Voigt functions given in \eqref{eq:ReVoigt} and \eqref{eq:ImVoigt}, respectively. The PDF of $Z_n$ is now stated in the following theorem:

\begin{theorem}
		\label{thm:PDFNoiseChanC}
		Let $Z_n \sim \StabDist(0,c_{\sysC},\tfrac{1}{2},\beta_{\sysC})$. Then the PDF of $Z_n$ is given by:
		\begin{align}
		\label{eq:noisePDFchanC}
		f_{Z_n}(z_n) = \begin{cases} 
		\frac{1}{c_{\sysC}} G_+\left( \tfrac{z_n}{c_{\sysC}}, \beta_{\sysC}  \right), & z_n > 0 \\
		\frac{2(1-\beta^2)}{c_{\sysC}\pi(1+\beta^2)^2}, & z_n = 0  \\
		\frac{1}{c_{\sysC}} G_-\left( \tfrac{z_n}{c_{\sysC}}, \beta_{\sysC}  \right), & z_n < 0 
		\end{cases}. 
		\end{align}
\end{theorem}
\begin{IEEEproof}
 The proof is provided in Appendix \ref{annex:ProofOfPDFchanC}.
\end{IEEEproof}

Again, we do not provide an expression for the CDF of the noise terms $F_{Z_n}(z_n)$, instead we note that it can be numerically calculated using the methods of \cite[Sec. 3]{nol97}. 
Figs. \ref{fig:stablePDF} and \ref{fig:stableCDF} depict the PDF and CDF for the standardized noise term $Z_n$ with $c_{\sysC} =1$ and four different values for $\beta_{\sysC}$. Note that for $\beta_{\sysC} = 0$ the distribution and the density functions of the noise $Z_n$ in System C are the same as the noise $L_n$ in system B. This is due to the diffusion coefficient of type-$a$  and type-$b$ particles being equal, which means that both particle types have the same random propagation delay characteristics.  

\revised{
\begin{rem}
	Moving from a 1-D space to a 3-D space, there is a probability that the particle will never arrive at the receiver (see \cite{yilmaz20143dChannelCF}). For systems B and C, both particles must arrive for error free communication. Conditioned on the event that the two particles arrive, the noise distribution will be the same as the 1-D case. Therefore, in 3-D, the noise distribution will be scaled by the probability of the event that both particles arrive.
\end{rem}
}

\section{Optimal Detection in Binary DBMT Systems}
\label{sec:PBE}
In this section, we consider equiprobable binary transmission over the three different DBMT systems. Using the noise models developed in the previous section, we characterize the optimal detection rule for each modulation.

\subsection{System A}
For system A, \revised{we} assume that the transmission symbols are $T_x \in \{0,\Delta\}$, where $\Delta>0$. 
Using Property \ref{prop:standStable}, we write the distribution of the output probability, conditioned on the input, in terms of the standard \levy distribution $\tilde{T}_n \sim \LevyDist(0,1)$ as follows:
\begin{align}
f_{T_y|T_x}(t_y|T_x=0) &= f_{T_n} (t_y) = \frac{f_{\tilde{T}_n}(t_y/c_{\sysA})}{c_{\sysA}} , \\
f_{T_y|T_x}(t_y|T_x=\Delta) &= f_{T_n} (t_y-\Delta) = \frac{f_{\tilde{T}_n}\big((t_y-\Delta)/c_{\sysA}\big)}{c_{\sysA}}. 
\end{align}
\begin{figure*}
	\begin{center}
		\includegraphics[width=0.7\textwidth]{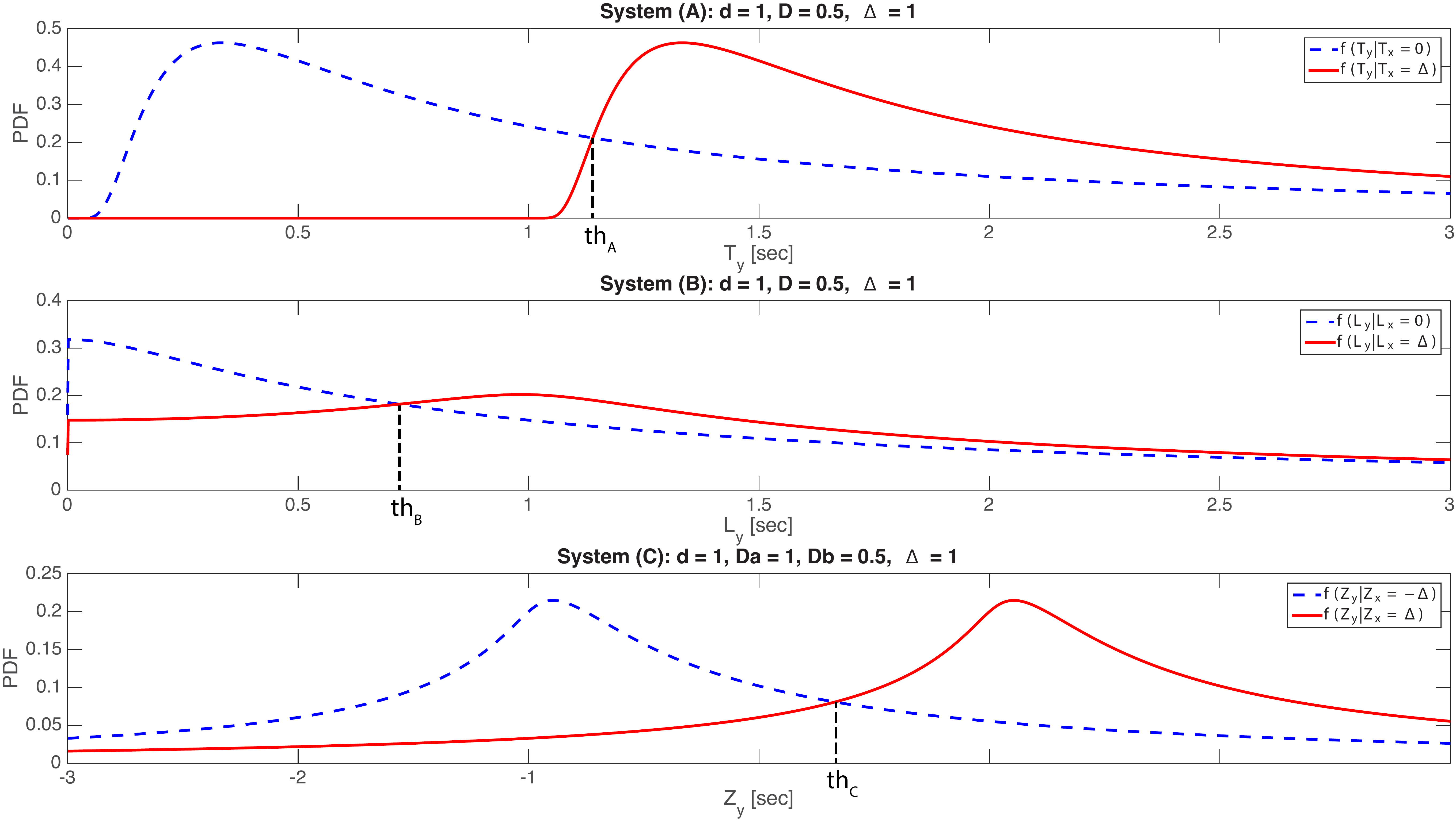}
	\end{center}
	\caption{\label{fig:optiThreshold} The ML optimal decision threshold for the three systems.}
\end{figure*}

As the two transmitted symbols are equiprobable, the detector that minimizes the probability of error is the maximum likelihood (ML) detector. 
In this work we assume both the 0-bit and the 1-bit are equiprobable and apply the ML detector. In this case, the likelihood ratio is given by:
\begin{align}
\Lambda_A (t_y) = \frac{f_{T_y|T_x}(t_y|T_x=0)}{f_{T_y|T_x}(t_y|T_x=\Delta)}, \label{eq:LR_A}
\end{align}

\noindent and optimal detection can be done by a comparison of the log likelihood ratio (LLR) to zero, i.e.,
\begin{align}
\log(\Lambda_A (t_y) ) \begin{matrix} T_x = 0 \\ \gtrless \\ T_x = \Delta \end{matrix} 0.
\end{align} 
Note that the proof of the existence of the optimal threshold value is straightforward using the fact that stable distributions are unimodal \cite[Theorem 2.7.6]{zol86-book}, and that for the noise term $T_n$ the mode is at $c/3$. Therefore, there exists a threshold $\Delta<\thr_{\sysA}\leq c/3+\Delta$, such that $\Lambda_A (t)>1$ for $t<\thr_{\sysA}$  and  $\Lambda_A (t)\leq1$ for $t\geq\thr_{\sysA}$ \cite{mur16Globecom,mur17}. 
The top plot in Figure \ref{fig:optiThreshold} shows the optimal threshold for the case when $\Delta=1$, the distance is $d=1$ and the diffusion coefficient is $D=0.5$.

The probability of error for system A is now given by:
\begin{align}
P_e^A &= P(T_x=0) \Pr(t_y>\thr_{\sysA} | T_x = 0) \nonumber \\
&\qquad +P(T_x=\Delta) \Pr(t_y\leq\thr_{\sysA} | T_x= \Delta), \\
&= 0.5 \Pr(t_n>\thr_{\sysA})+0.5 \Pr(t_n\leq\thr_{\sysA}-\Delta) \\
&= 0.5[1-F_{\tilde{T}_n}(\tfrac{\thr_{\sysA}}{c_{\sysA}})+F_{\tilde{T}_n}(\tfrac{\thr_{\sysA}-\Delta}{c_{\sysA}})],
\end{align} 
where $F_{\tilde{T}_n}(t)$ is the CDF of a standard L\'evy RV. 


\subsection{System B}
For system B we assume that the input is $L_x\in\{0,\Delta\}$, where $L_x=0$ represents two particles released simultaneously, while $L_x=\Delta$ represents two particles released $\Delta$ seconds apart. Let $\tilde{L}_n \sim \StabDist(0,1,\tfrac{1}{2},0)$ be the standard form of the noise term in \eqref{eq:timingChB}. 
The PDF of the output $L_y$, given the input $L_x$, is provided in the following proposition:
\begin{proposition}
	The system output $L_y$, given the system input $L_x$, has the PDF: 
%
\begin{align}
f_{L_y|L_x}(\ell_y|L_x=0) &= 
\begin{cases}
\frac{2f_{\tilde{L}_n}\big(\tfrac{\ell_y}{c_{\sysB}}\big)}{c_{\sysB}} & \ell_y>0 \\
\frac{2}{(c_{\sysB}\pi)} & \ell_y=0 \\
0 & \ell_y<0
\end{cases}, \label{eq:LyGLx0}\\
f_{L_y|L_x}(\ell_y|L_x=\Delta) &= 
\begin{cases}
\frac{f_{\tilde{L}_n}\big( \tfrac{\ell_y-\Delta}{c_{\sysB}}\big)+f_{\tilde{L}_n}\big( \tfrac{-\ell_y-\Delta}{c_{\sysB}}\big)}{c_{\sysB}} & \ell_y>0 \\
\frac{f_{\tilde{L}_n}\big( \tfrac{\Delta}{c_{\sysB}}\big)}{c_{\sysB}} & \ell_y=0 \\
0 & \ell_y<0
\end{cases}. \label{eq:LyGLxD} 
\end{align}

\end{proposition}

%

\begin{IEEEproof}
	It is clear from the system definition that when $L_y<0$ the PDF is 0 (i.e. the time between two arrival times is not negative). When $L_y=0$, we have $f_{L_y|L_x}(0|L_x=0)=f_{L_n}(0)$, and $f_{L_y|L_x}(0|L_x=\Delta)=f_{L_n}(\Delta)$.  To derive the PDF value for $L_y>0$, we use the fact that the CDF of $L_y$ given $L_x=x\geq0$ can be obtained from the CDF of $\tilde{L}_n$ as
	\begin{align*}
	F_{L_y|L_x}(\ell_y|L_x=x) &= \Pr(L_y \leq \ell_y|L_x=x) \\
	&= \Pr(|x+L_n| \leq \ell_y) \\
	&= \Pr(-\ell_y \leq x+L_n \leq \ell_y) \\
	&= \Pr(\tfrac{-\ell_y-x}{c_{\sysB}} \leq \tilde{L}_n \leq \tfrac{\ell_y-x}{c_{\sysB}} ) \\
	&= F_{\tilde{L}_n}\big(\tfrac{\ell_y-x}{c_{\sysB}}\big)-F_{\tilde{L}_n}\big(\tfrac{-\ell_y-x}{c_{\sysB}}\big).
	\end{align*}
	By differentiating with respect to $\ell_y$, and setting $x=0$ and $x=\Delta$, we obtain \eqref{eq:LyGLx0} and \eqref{eq:LyGLxD}, respectively. 
\end{IEEEproof}

Similarly to \eqref{eq:LR_A}, the likelihood ratio for the ML detector for system B is given by:
\begin{align}
\Lambda_B (\ell_y) = \frac{f_{L_y|L_x}(\ell_y|L_x=0)}{f_{L_y|L_x}(\ell_y|L_x=\Delta)}.
\end{align}

\noindent The following theorem states that, just like in the case of system A, the ML detector can be implemented by comparing $\log (\Lambda_B (\ell_y))$ to zero:
\begin{theorem}
	\label{thrm:ChanBthrExistance}
	There exists a fixed threshold $\thr_{\sysB} >\tfrac{\Delta}{2}$ such that the ML detector in the case of system B is given by:
	\begin{align}
		\log(\Lambda_B (\ell_y) ) \begin{matrix} T_x = 0 \\ \gtrless \\ T_x = \Delta \end{matrix} 0.
	\end{align} 
\end{theorem}
\begin{IEEEproof}
	The proof is provided in Appendix \ref{annex:ProofOfExistanceOfThreshold}.
\end{IEEEproof}
The middle plot in Fig. \ref{fig:optiThreshold} depicts the optimal threshold for the case when $\Delta=1$, the distance is $d=1$ and the diffusion coefficient is $D=0.5$. Since the closed-form expression for the CDF of the noise term is unknown, this threshold is calculated numerically. 
Finally, the probability of error for binary communication over system B is given by:
\begin{align}
P_e^B &= P(L_x=0) \Pr(L_y>\thr_{\sysB} | L_x = 0) \nonumber \\
&\qquad +P(L_x=\Delta) \Pr(L_y\leq\thr_{\sysB} | L_x= \Delta), \nonumber \\
&= 0.5(\Pr(L_n>\thr_{\sysB})+\Pr(L_n\leq -\thr_{\sysB}) )\nonumber \\
&\qquad +0.5\Pr(-\thr_{\sysB}-\Delta \leq L_n\leq \thr_{\sysB}-\Delta), \nonumber \\
&= 0.5(\Pr(\tilde{L}_n>\tfrac{\thr_{\sysB}}{c_{\sysB}})+\Pr(\tilde{L}_n\leq -\tfrac{\thr_{\sysB}}{c_{\sysB}}) )\nonumber \\
&\qquad +0.5\Pr(\tfrac{-\thr_{\sysB}-\Delta}{c_{\sysB}} \leq \tilde{L}_n\leq \tfrac{\thr_{\sysB}-\Delta}{c_{\sysB}}), \nonumber \\
&= F_{\tilde{L}_n}\big(\tfrac{\thr_{\sysB}}{c_{\sysB}}\big) \mspace{-2mu} + \mspace{-2mu} 0.5\big(F_{\tilde{L}_n}\big(\tfrac{\thr_{\sysB}-\Delta}{c_{\sysB}}\big) \mspace{-2mu} - \mspace{-2mu} F_{\tilde{L}_n}\big(\tfrac{\thr_{\sysB}+\Delta}{c_{\sysB}}\big)\big).
\end{align} 
Thus, similarly to the case of system A, the probability of error can be calculated using the standard form of the noise term.

\subsection{System C}
Recall that for system C the two particles are {\em distinguishable}, and $Z_x=T_{x_b}-T_{x_a}$ is the time interval between the releases of particles $b$ and $a$. Here, we assume information is encoded in the {\em order} of release. The input $Z_x\in\{-\Delta,\Delta\}$ is now given by:
\begin{align}
	Z_x = \begin{cases} \Delta, & T_{x_a} = 0, T_{x_b} = \Delta \\
											-\Delta, & T_{x_b} = 0, T_{x_a} = \Delta \end{cases}.
\end{align}

\noindent Note that similarly to systems A and B, the information is encoded over the time period $\Delta$.

Let $\tilde{Z}_n \sim (0,1,\tfrac{1}{2},\beta_{\sysC})$ be the standard form of the noise term in \eqref{eq:timingChC}. Then the PDF of the output given the input is given by
\begin{align}
f_{Z_y|Z_x}(z_y|Z_x=-\Delta) &= \frac{f_{\tilde{Z}_n}\big(\tfrac{z_y+\Delta}{c_{\sysC}}\big)}{c_{\sysC}}\\
f_{Z_y|Z_x}(z_y|Z_x=\Delta) &= \frac{f_{\tilde{Z}_n}\big(\tfrac{z_y-\Delta}{c_{\sysC}}\big)}{c_{\sysC}}. 
\end{align}
Again, to minimize the probability of error at the receiver, the ML detector is used.
Let $\thr_{\sysC}$ be the optimal ML detection threshold for this system. It is easy to see that this threshold exists for system C since stable distributions are unimodal and the two PDFs are shifted versions of each other. The bottom plot in Figure \ref{fig:optiThreshold} shows the optimal threshold for the case when $\Delta=1$, the distance is $d=1$ and the diffusion coefficients are $D_a=1$ and $D_b=0.5$.
The probability of error is now given by: 
\begin{align}
\label{eq:PeC}
P_e^C &= P(Z_x=-\Delta) \Pr(z_y>\thr_{\sysC} | Z_x = -\Delta) \nonumber \\
&\qquad +P(Z_x=\Delta) \Pr(z_y\leq\thr_{\sysC} | Z_x= \Delta), \nonumber \\
&= 0.5 \Pr(z_n>\thr_{\sysC}+\Delta)+0.5 \Pr(z_n\leq\thr_{\sysC}-\Delta) \nonumber \\
&= 0.5[1-F_{\tilde{Z}_n}(\thr_{\sysC}+\Delta)+F_{\tilde{Z}_n}(\thr_{\sysC}-\Delta)],
\end{align}
which can be calculated using the CDF of the standard form of the noise term. 

\section{Geometric Power and G-SNR} \label{sec:SNR}

We first note that all stable distributions, apart from the case $\alpha = 2$, have infinite variance, and all stable distributions with $\alpha \le 1$ also have infinite mean. In fact, this statement can be generalized to moments of order $p \le \alpha$, see \cite{gon06}. Therefore, the conventional notion of power, which is based on the variance of a signal, is not informative in the case of stable RVs with $\alpha < 2$ as, regardless of the specific distribution, the conventional power is infinity. 
In this section we use a more generalized definition of power, the {\em geometric power}, as proposed in \cite[Section III]{gon06}. This definition uses zero-order statistics, i.e., it is based on logarithmic ``moments" of the form $\mathbb{E}[\log|N|]$.
%
%
\begin{definition}[Geometric Power]
	The geometric power of the RV $N$ is given by:
	\begin{align}
	S_0(N) \triangleq e^{\mathbb{E} [\log|N|]}.
	\end{align}  
\end{definition}
In the following we use the terms {\em noise power} and the geometric power of the noise interchangeably.\footnote{Note that the definition of geometric power/SNR we introduce here is different from the one widely used in RF communications.}  

In \cite[Prop. 1]{gon06}, an expression for the geometric power of a symmetric stable distribution is presented. Property \ref{prop:sym} implies that symmetric stable distributions are in fact $\StabDist (0,c,\alpha,0)$. 
This expression can therefore be used to calculate the geometric power of the noise term $L_N$ in system B. Yet, this expression is not applicable for the noise terms of systems A and C in which $\beta \neq 0$. The following theorem characterizes the geometric power of almost all stable distributions:
\begin{theorem}
	\label{thrm:GpowerStable}
	Let $N \sim \StabDist (0,c,\alpha, \beta)$, where $\alpha \neq 1$, or $\alpha = 1$ and $\beta=0$. Then, the geometric power of $N$ is given by:
	\begin{align}
		S_0(N) = c G_\gamma^{(1/\alpha-1)} \big(1+\beta^2\tan^2(\tfrac{\pi \alpha}{2})\big)^{1/(2\alpha)}, 
	\end{align}
	where $G_\gamma = e^\gamma$, and $\gamma \approx 0.5772$ is the Euler's constant \cite[Ch. 5.2]{nist10}.
\end{theorem}
\begin{IEEEproof}
	The proof is provided in Appendix \ref{annex:ProofOfGeomPower}.
\end{IEEEproof}

\begin{rem}
	For the systems considered in this paper, since $\alpha = \frac{1}{2}$, the noise power simplifies to: 
	\begin{align}
		S_0(N) = cG_\gamma\big(1+\beta^2\big).
	\end{align}
	Note that in this case, the noise power increases with respect to $\beta$ (the degree of skewness) and $c$ (the scale parameter).
\end{rem}

We now define the geometric SNR (G-SNR) as in \cite[Section III]{gon06}:
\begin{definition}[Geometric Signal-to-Noise Ratio]
	\label{def:GSNR}
	Let $X$ be the input signal in an additive-noise channel with a random noise $N$. Then the G-SNR is defined as: 
	\begin{align}
	\label{eq:GSNRdef}
	\gsnr \triangleq \frac{1}{2 G_\gamma }\bigg(\frac{X_{\max} - X_{\min}}{S_0(N)}\bigg)^2,
	\end{align}  
	where $X_{\max}$ and $X_{\min}$ are the maximum and minimum admissible values for the channel input $X$. The normalizing term $\frac{1}{2 G_\gamma}$ is used to ensure that the G-SNR corresponds to the standard SNR in the case of an additive Gaussian noise channel.
\end{definition}

Using this definition and Theorem \ref{thrm:GpowerStable}, the G-SNR for systems A and C is defined as follows:
	\begin{align}
	\gsnr_A &= \frac{1}{2 G_\gamma }\bigg(\frac{\Delta}{2 c_{\sysA} G_\gamma }\bigg)^2, \label{eq:GSNRa}\\
	\gsnr_C &= \frac{1}{2 G_\gamma }\bigg(\frac{2\Delta}{ c_{\sysC} G_\gamma(1+\beta_{\sysC}^2) }\bigg)^2. \label{eq:GSNRc}
	\end{align} 	
	
\begin{rem} \label{rem:absValImpact}
	Note that system B involves an absolute value operation, thus, the G-SNR of system B cannot be obtained based on the techniques used to derive the G-SNR for systems A and C. Since the absolute value operation can only degrade the detection performance, calculating the G-SNR of the system $L_y = L_x + L_n$ can serve as an upper bound on the G-SNR of system B. This upper bound is given by:
	\begin{align}
		\gsnr_B \le \gsnr_B^{\text{ub}} = \frac{1}{2 G_\gamma }\bigg(\frac{\Delta}{ c_{\sysB} G_\gamma }\bigg)^2. \label{eq:GSNRb}
	\end{align}
	
\noindent This implies that the BER of the ML detector for system B is higher than the BER of the ML detector for the system $L_y = L_x + L_n$, as indicated in Section \ref{sec:numEval}.

\end{rem}

\begin{rem}
	When the diffusion coefficient and the distance between the transmitter and the receiver are the same, the G-SNR of system A is four times larger than the G-SNR of system B since $c_{\sysB}=4c_{\sysA}$. 
	This implies that on top of the fact that two information particles are released in system B while only a single particle is released in system A, the gain from synchronization is a factor of $\frac{1}{4}$ in the noise geometric power.
\end{rem}

 \begin{figure}
 	\begin{center}
 		\includegraphics[width=\columnwidth]{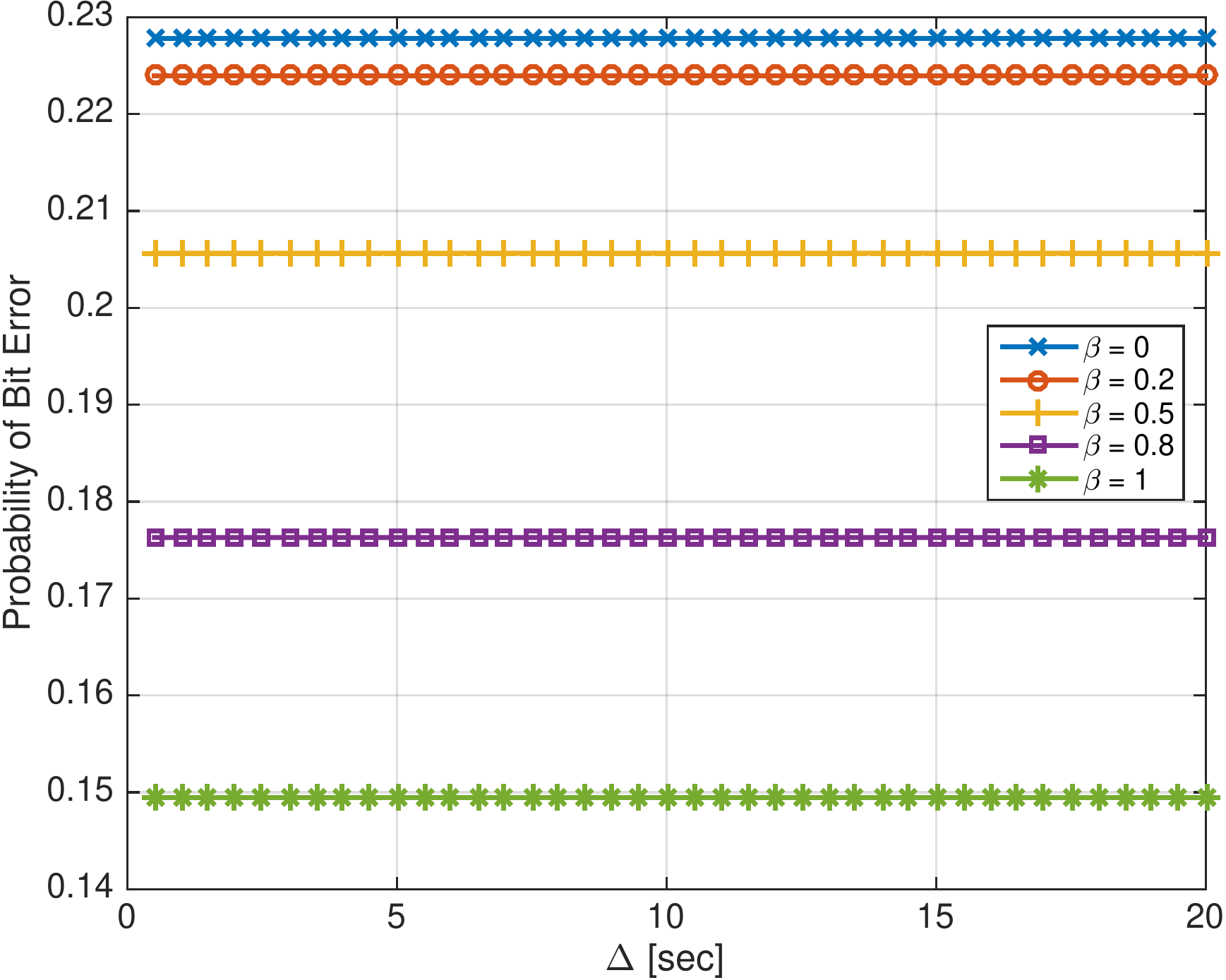}
 	\end{center}
 	\caption{\label{fig:proofSNR1} This plot shows that for a constant G-SNR, the BER is constant. For each point, the parameter $c$ of the noise distribution is calculated using the corresponding value for $\Delta$ such that the G-SNR$=1$. }
 \end{figure}

\begin{rem}
\label{rem:effectBeta}
For system C let $r=D_a/D_b$ be the ratio of the diffusion coefficient of the two information particles. Then the noise parameters can be written as $c_{\sysC}= \frac{d^2(\sqrt{r}+1)^2}{2rD_b}$ and $\beta_{\sysC} = \frac{\sqrt{r}-1}{\sqrt{r}+1}$. Next, assume that the diffusion coefficient $D_b$ is fixed, and the diffusion coefficient of $D_a$ can be changed. In this case the noise geometric power is proportional to $\frac{1}{r}$, which decreases as $r$ increases. This also implies that the  G-SNR increases with $r$. From the expression for $\beta_{\sysC}$ and $c_{\sysC}$ we observe that $\beta_c \rightarrow 1$ and $c_{\sysC} \rightarrow \tfrac{d^2}{2D_b}$, when $r \rightarrow \infty$. Thus, in this case, system C reduces to system A, while no synchronization is required between the transmitter and the receiver.
Yet, this comes at a cost of using two different information particles.  Note that this cost is captured in the G-SNR expression since the geometric power of the transmitted signal in system C is four times that of systems A and B, which can result in as much as 4 times improvement in G-SNR.
\end{rem}

\section{Numerical Evaluation}
\label{sec:numEval}
We start this section by evaluating the affects of the G-SNR, provided in \eqref{eq:GSNRdef}, on the BER performance of the three modulation schemes. In the additive white Gaussian noise channel the BER of the ML detector is only a function of SNR, namely, for a fixed SNR, the individual values of the signal power and the noise power do not affect the BER.
To evaluate if this property also holds for the three MT systems, we consider system C which can be specialized to both systems A and B using different values of the parameter $\beta_{\sysC}$, see \eqref{eq:GSNRa}--\eqref{eq:GSNRc}. Thus, we evaluate if a constant BER is observed for a fixed value of G-SNR. 

\begin{figure}
  	\begin{center}
  		\includegraphics[width=\columnwidth]{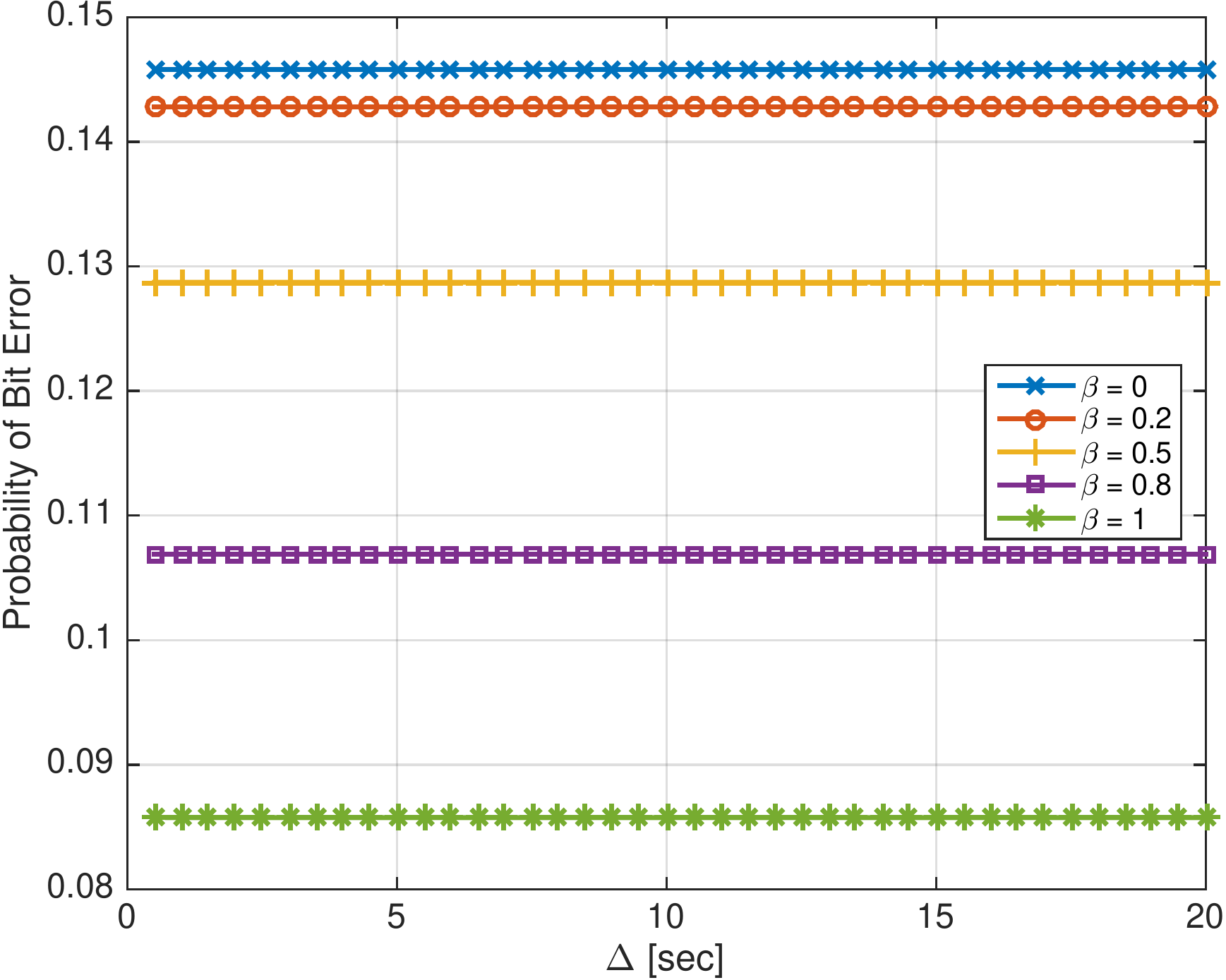}
  	\end{center}
  	\caption{\label{fig:proofSNR10} This plot shows that for a constant G-SNR, the BER is constant. For each point, the parameter $c$ of the noise distribution is calculated using the corresponding value for $\Delta$ such that the G-SNR$=10$.}
  \end{figure}

Figs. \ref{fig:proofSNR1} and \ref{fig:proofSNR10} depict BER versus $\Delta$ for two values of G-SNR: 1 and 10. In these plots, the x-axis corresponds to the values of $\Delta$. 
For each point in the plot, the value of the noise parameter $c_{\sysC}$ is calculated such that G-SNR is either 1 (Fig. \ref{fig:proofSNR1}) or 10 (Fig. \ref{fig:proofSNR10}). The BER is then numerically calculated using these values based on \eqref{eq:PeC}. It can clearly be observed that the BER is constant for a given G-SNR {\em regardless} of the value of $\Delta$ and $c_{\sysC}$. It can further be observed that the BER decreases as $\beta_{\sysC} \rightarrow 1$, which is in agreement with Remark \ref{rem:effectBeta}. 

\begin{figure}
  	\begin{center}
  		\includegraphics[width=\columnwidth]{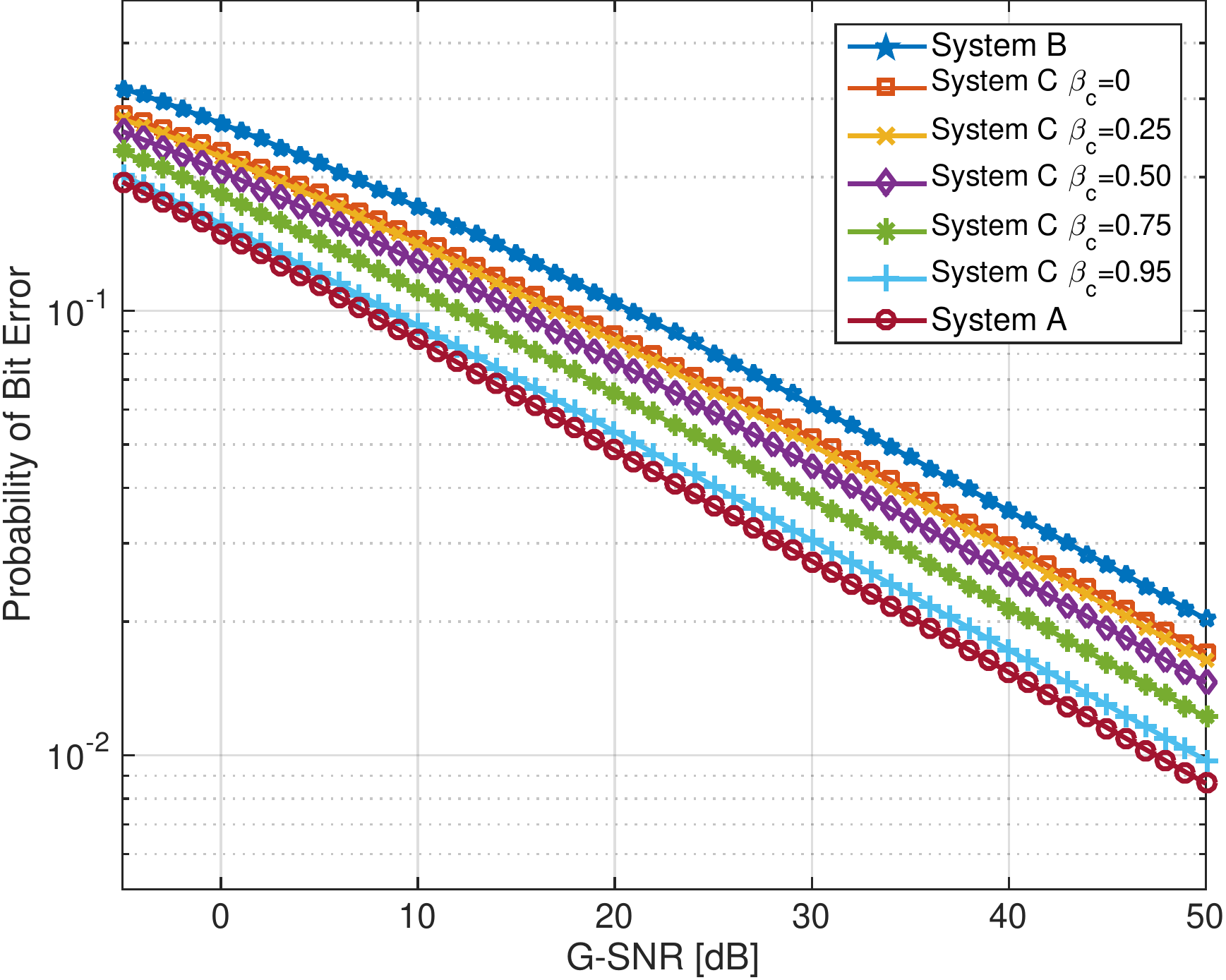}
  	\end{center}
  	\caption{\label{fig:PBEvsGSNR} BER versus G-SNR in dB for each modulation scheme.}
  \end{figure}

Fig. \ref{fig:PBEvsGSNR} depicts the BER versus G-SNR for the different modulation techniques. For system C, five different values of $\beta_{\sysC} =0,0.25,0.5,0.75,0.95$ are considered. The asynchronous scheme in system B with indistinguishable particles achieves the highest BER, while system A, which assumes perfect synchronization, achieves the lowest BER. The gap between these can be thought of as the cost of having no synchronization. Note that in system A, a single particle is released, while in system B two particles are released. 

For system C it can be observed that by using two distinguishable particles, the BER improves compared to system B. Note that when $\beta_{\sysC} = 0$ the noise distribution is the same as that in system B. In this case, when the dispersion parameter $c$ is the same for both systems, the G-SNR of system C is four times larger than $\gsnr_B^{\text{ub}}$ in \eqref{eq:GSNRb}. Yet, Fig. \ref{fig:PBEvsGSNR} indicates that even for $\beta_{\sysC} = 0$ the BER of system C is lower than the BER of system B. This demonstrates the destructive effect of the absolute value operation as indicated in Remark \ref{rem:absValImpact}.
Finally, we observe that as $\beta_{\sysC}$ increases the BER of system C decreases, while when $\beta_{\sysC} \rightarrow 1$ the BER of system C approaches the BER of system A. In this case, asynchronous communication is possible with the same BER performance as synchronized communication at the cost of using two distinguishable particles. 

We conclude the numerical evaluations with a case study. We consider a DBMT system where the distance between the transmitter and the receiver is $d = 20$ $\mu$m. Assume that the receiver is capable of detecting insulin molecules, which has a diffusion coefficient of $D_I = 150$ $\mu$m$^2$/s \cite{far16ST}. From these values the noise parameters $c_{\sysA}$ and $c_{\sysB}$ can be calculated for the modulation techniques represented by systems A and B. For system C, we consider six different particles as candidates for the second distinguishable particle. These particles are assumed to have diffusion coefficients ranging from 30 $\mu$m$^2$/s (e.g., diffusion coefficient of DNA) to 930 $\mu$m$^2$/s (e.g., diffusion coefficient of glycerol).

\begin{figure}
	\begin{center}
		\includegraphics[width=0.9\columnwidth]{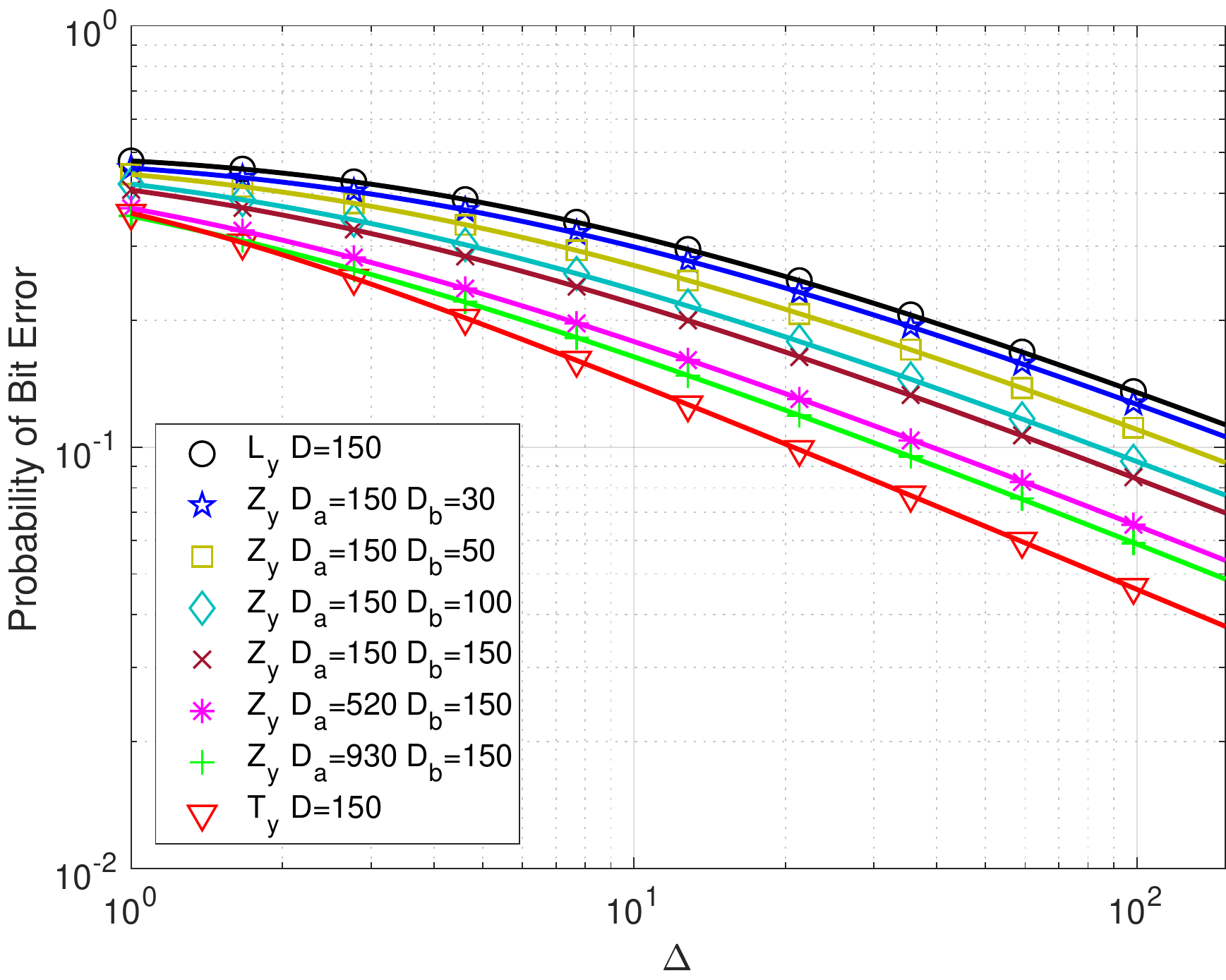}
	\end{center}
	\caption{\label{fig:PBEvsDelta} BER versus $\Delta$ under different modulation schemes. \revised{The line plots are based on the theoretical results derived in the paper, and the point plots are the corresponding Monte Carlo simulation results.}}
\end{figure}

Fig.~\ref{fig:PBEvsDelta} depicts the results. \revised{In order to further evaluate the correctness of the analytical results, we also perform Monte Carlo simulations. It can be observed that the theoretical results (line plots) match perfectly with the results of the Monte Carlo simulations (point plots).} The asynchronous modulation scheme in system B with indistinguishable particles has the highest BER. Note that even the modulation scheme in system C where the diffusion coefficient of the second particle is one fifth of the diffusion coefficient of the particles used in system B (i.e. 30 $\mu$m$^2$/s) has lower BER. For the modulation technique in system C, as the diffusion coefficient of the second particle increases, the BER decreases. The modulation in system A achieves the best BER performance, and this shows that transmitter-receiver synchronization could have a considerable effect on BER.  Table \ref{tab:PBE} quantifies the BER for each case.

\small 
\begin{table}[h]
    \caption{\revised{The BER of different modulations in the case study for different values of $\Delta$. For the modulation scheme in system C the term in parenthesis is the diffusion coefficient of the second particle.} \label{tab:PBE}}
    \vspace{-0.5cm}
	\begin{center}

		\begin{tabular}[t]{|c|c|c|c|c|c|}
			\hline
			$\Delta$ & 1  & 25 & 50 & 75 & 100  \\
			\hline
			\hline
			System A   & 0.3590  &  0.0912  &  0.0648  &  0.0530  &  0.0460  \\
			\hline
			System B  & 0.4778  &  0.2346  &  0.1799  &  0.1523 &   0.1348  \\
			\hline
			System C ($D=30$)  &  0.4592  &  0.2202  &  0.1687  &  0.1428  &  0.1263 \\
			\hline
			System C ($D=150$)  &  0.4073  &  0.1535  &  0.1145  &  0.0957 &  0.0841 \\
			\hline
			System C ($D=930$)  &  0.3533  &  0.1109  &  0.0812  &  0.0674  &  0.0589  \\
			\hline
		\end{tabular}
	\end{center}

\end{table}  
\normalsize

\revised{Finally, recall that the above results are derived under the assumption of an ISI-free channel (see Assumption \ref{assmp:noISI}). This can be achieved by properly spacing the transmissions leading to relatively large symbol durations. We now discuss how large this duration should be for the different \nariman{considered} systems.   
    Let $T_{\mathrm{symbol}}$ denote the spacing between consecutive transmissions, and $0 < p_{\mathrm{clean}} < 1$. Further, let $T_{\mathrm{last}}$ denote the last particle arrival time {\em calculated over the particles released in the current channel use}.  
    We propose to choose $T_{\mathrm{symbol}}$ such that:
    \begin{align*}
    	\Pr\{ T_{\mathrm{last}} \nariman{\le} 
        T_{\mathrm{symbol}} \} \nariman{=} 
        p_{\mathrm{clean}}.
    \end{align*}
    
    \noindent Hence, if $p_{\mathrm{clean}}$ is a (fixed) value close to 1, then with high probability all the released particles arrive \nariman{at} 
    the receiver \nariman{before} 
    $T_{\mathrm{symbol}}$, implying that after this idle duration the channel can be used for another transmission. Moreover, the symbol duration $T_{\mathrm{symbol}}$ can be found from the CDF of $T_{\mathrm{last}}$ as a function of $p_{\mathrm{clean}}$. Note that the CDF of $T_{\mathrm{last}}$ is different for \nariman{each} 
    system \nariman{considered}. Yet, Assumption \ref{assmp:indep} implies that for all three systems the CDF of $T_{\mathrm{last}}$ is the product of the CDFs of the propagation time of the individual particles (in the case of System A it is simply the CDF of the L\'evy distribution).
    
    Fig. \ref{fig:SymbolDurationVsDelta} depicts the calculated $T_{\mathrm{symbol}}$ for the different systems with different diffusion coefficients. $p_{\mathrm{clean}}$ was set to $0.99$. It can be observed that due to the heavy tails of the propagation density, the symbol duration is almost constant as a function of $\Delta$, and is much larger than $\Delta$. It can further be observed that the order of the curves in Fig. \ref{fig:SymbolDurationVsDelta} is almost identical to the order of the curves in Fig. \ref{fig:PBEvsDelta}, where the exception is the curve corresponding to System B which requires almost the same $T_{\mathrm{symbol}}$ as System C with the same diffusion coefficients. 
    Note that the modulation in system A requires the smallest $T_{\mathrm{symbol}}$, thus, together with the results of Fig. \ref{fig:PBEvsDelta}, \nariman{we conclude} 
    that the performance gains obtained from transmitter-receiver synchronization are significant. 
    
    \begin{figure}
	\begin{center}
		\includegraphics[width=0.975\columnwidth]{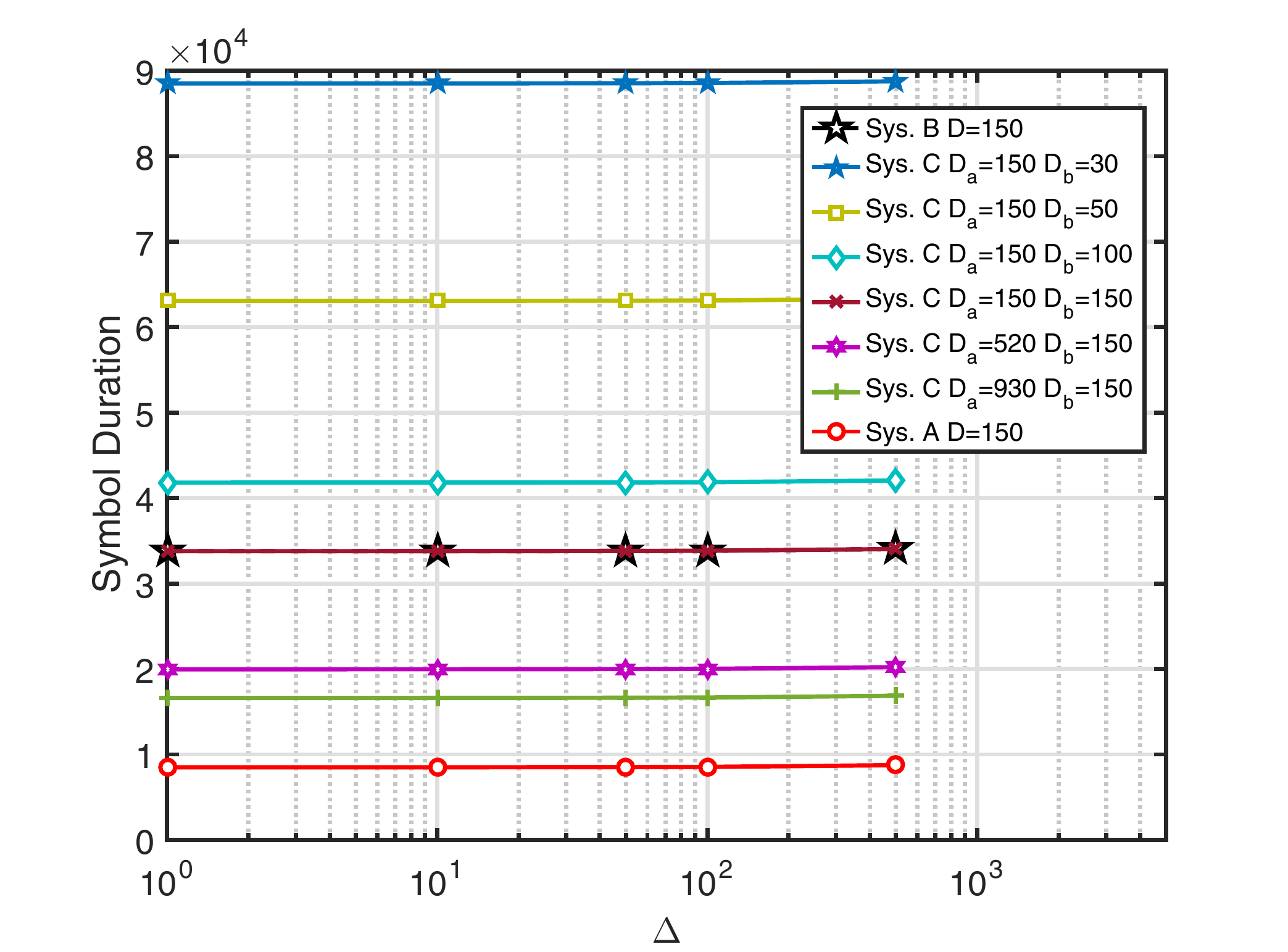}
	\end{center}
	\caption{\label{fig:SymbolDurationVsDelta} $T_{\mathrm{symbol}}$ versus $\Delta$ under different modulation schemes, for $p_{\mathrm{clean}}=0.99$.}
\end{figure}

    }

\section{Conclusions and Future Work}
\label{sec:conc}
In this work, we considered two new asynchronous timing-based modulation techniques based on the time between release of two similar information particles, and the time between release of two different information particles. For evaluation, we compared the performance of these systems to the synchronized modulation based on the time of release of information particles. We showed that the three modulation techniques can be modeled as systems with an additive noise term, where for diffusion-based propagation, the noise terms are stable distributed. For the asynchronous systems, we derived the PDF of the additive noise in terms of the Voigt functions, which can be calculated efficiently and in some special cases be approximated using elementary functions. 
Using these PDFs we then characterized the ML detectors for each system. 
Since stable distributions, with the exception of the Gaussian distribution, have infinite variance, we used geometric power as a measure of strength of the noise. Using this approach, we derived the G-SNR for each modulation scheme for comparison. 
Numerical evaluations show that for a constant G-SNR the BER is constant. Therefore, the G-SNR in DBMT channels plays a similar role as the SNR in the additive Gaussian noise channels.
Finally, we showed that, as expected, synchronization has a considerable effect on BER, where the first modulation scheme achieves the lowest BER. Moreover, we showed that it is possible to achieve a similar BER asynchronously if two distinguishable particles are used per bit.

As part of future work, we will explore extending the results to the case where multiple information particles are released simultaneously instead of one. Note that some of our current ongoing work has shown that simultaneously releasing multiple particles can improve the performance of the first system significantly \cite{far17TIT,mur17}. We would like to extend these results to the asynchronous systems presented in this paper using order statistics.

 \appendices
  \section{Proof of Theorem \ref{thm:PDFNoiseChanB}} \label{annex:ProofOfPDFchanB}
  We use Property \ref{prop:standStable}, and find an expression for the standardized distribution with $c_{\sysB}=1$. Then the PDF for any value of $c_{\sysB}$ can be calculated using \eqref{eq:standPDFConv}. 
	Let $X \sim \StabDist (0,1,\tfrac{1}{2},0)$ be a standardized stable RV with parameters $\alpha = \tfrac{1}{2}$ and $\beta=0$.  Then the PDF of $X$ is given by \cite[Eq. (7.1)]{hol73}:
  \begin{align}
  \label{eq:alphaHalf}
  f(x;1/2,\beta) = \Re \left\{ \frac{z}{\pi x} [ \sqrt{\pi} e^{-z^2} - 2 j D(z)] \right\},
  \end{align}
  where
  \begin{align}
  D(z) = e^{-z^2} \int_0^z e^{t^2} dt
  \end{align}
  is the Dawson's Integral \cite[Eq. (7.2.5)]{nist10}, and 
  \begin{align}
  \label{eq:z}
  z = \frac{1+\beta - j(1-\beta)}{2\sqrt{2x}}.
  \end{align}
  
	\noindent It is possible to rewrite (\ref{eq:alphaHalf}) in terms of the complex error function, also known as Faddeeva function or the Kramp function \cite[Eq. (7.2.3)]{nist10}:
  \begin{align}
  w(z) = e^{-z^2} \left( 1+ \frac{2 j}{\sqrt{\pi}} \int_0^z e^{t^2} dt \right) = e^{-z^2} \erfc(-jz).
  \end{align} 
  Using \cite[Eq. (7.5.1)]{nist10}:
  \begin{align}
  \label{eq:dawsonFadev}
  D(z) =0.5 j \sqrt{\pi}( e^{-z^2} -w(z)),
  \end{align}
  and the property $w(-z) = 2 e^{-z^2} - w(z)$ , we rewrite (\ref{eq:alphaHalf}) as:
  \begin{align}
  \label{eq:alphaHalf2}
  f(x;1/2,\beta) = \Re \left\{ \frac{z}{\sqrt{\pi} x} w(-z) \right\}.
  \end{align}
  One of the benefits of writing the PDF in terms of the complex error function is that there are a large body of works that considered calculating it numerically. Moreover, if $z = a+j b$, for $b>0$ the complex error function can be represented by its real and imaginary parts as \cite[Sec. 1]{abr11}:
  \begin{align}
  \label{eq:CompErToVoigt}
  w(a+ j b) = K(a,b)+j L(a,b),\quad b>0,
  \end{align}
  where $K(a,b)$ and $L(a,b)$ are the real and imaginary Voigt functions given in \eqref{eq:ReVoigt} and \eqref{eq:ImVoigt}, respectively.
  
  Using Property \ref{prop:sym}, the PDF of $X$ is symmetric. Hence, the density for $X\geq0$ is sufficient for characterizing the whole PDF. 
	Since $\beta = 0$, when $X>0$, we can write $z = p_{x} - j p_{x}$ where $p_{x}= 1/\sqrt{8x}$. Substituting \eqref{eq:CompErToVoigt} in \eqref{eq:alphaHalf2}, the density of $X$, when $X\geq0$, can be written as:
  \begin{align}
  \label{eq:fLn}
  f(x) = 
  \begin{cases}
  \frac{1}{\sqrt{8 \pi x^3}}\left[ K(-p_{x},p_{x})+L(-p_{x},p_{x})\right] & x>0 \\
  \frac{2}{\pi} & x=0
  \end{cases},
  \end{align}
  where the value for $x=0$ follows from \cite[Eq. (2.2.11)]{zol86-book}. Finally, the density for $X<0$ is obtained using symmetry. The proof is completed by applying \eqref{eq:standPDFConv}.

 \section{Proof of Theorem \ref{thm:PDFNoiseChanC}} \label{annex:ProofOfPDFchanC}
 We use Property \ref{prop:standStable}, and find an expression for the standardized distribution with $c_{\sysC}=1$. Thus, the PDF for any value of $c_{\sysC}$ can be calculated using \eqref{eq:standPDFConv}. 
Let $X \sim \StabDist (0,1,\tfrac{1}{2},\beta_{\sysC})$ be the standardized stable RV with parameters $\alpha = \tfrac{1}{2}$ and $\beta_{\sysC}$. 
Using \eqref{eq:alphaHalf2}, and recalling that  $\beta_{\sysC} = (\sqrt{D_a}-\sqrt{D_b})/(\sqrt{D_a}+\sqrt{D_b})$, we write (\ref{eq:z}) as $z=p_{x} - jq_{x}$ when $x>0$, where $p_{x}=(1+\beta_{\sysC})/(\sqrt{8|x|})$ and $q_{x}=(1-\beta_{\sysC})/(\sqrt{8|x|})$. Similarly, we write (\ref{eq:z}) as $z=-q_{x} - jp_{x}$ when $x<0$. Using (\ref{eq:alphaHalf2}) and the Voigt functions decomposition of the Faddeeva function \eqref{eq:CompErToVoigt}, the PDF of the standardized distribution is given by:
  \vspace{-0.15cm}
	\begin{align*}
  \label{eq:fZn}
  f(x; \beta_{\sysC}) = 
  \begin{cases}
  \frac{1}{\sqrt{8 \pi x^3}} \bigg[ (1+\beta_{\sysC}) K(-p_{x},q_{x})  &  \\
  \quad \quad  \quad ~~+(1-\beta_{\sysC})L(-p_{x},q_{x})\bigg], & x>0\\
  \frac{2(1-\beta^2)}{\pi(1+\beta^2)^2}, & x=0 \\
  \frac{1}{\sqrt{8 \pi |x|^3}}\bigg[(1-\beta_{\sysC})K(q_{x},p_{x}) &  \\
  \quad \quad \quad  ~~-(1+\beta_{\sysC}) L(q_{x},p_{x}) \bigg], & x<0
  \end{cases},
  \end{align*}
  
	\vspace{-0.1cm}
	\noindent where, again, the value for $x=0$ follows from \cite[Eq. (2.2.11)]{zol86-book}. The proof is completed by applying \eqref{eq:standPDFConv}.
 
\vspace{-0.2cm} 
\section{Proof of Theorem \ref{thrm:ChanBthrExistance}} \label{annex:ProofOfExistanceOfThreshold}

\vspace{-0.1cm}
 We first observe that for $\ell_y = 0$, $f_{L_y|L_x}(0 | 0)>f_{L_y|L_x}( 0| \Delta)$. This follows from the fact that stable distributions are unimodal, and the mode of the noise term $L_n$ is at $\ell =0$. Therefore, the threshold is located at $\thr_{\sysB}>0$, and we focus of the case where $\ell_y>0$. Note that in this case, due to the continuity and unimodality of stable distributions\cite[Theorem 2.7.6]{zol86-book}, both $f_{L_y|L_x}(\ell_y | L_x = 0)$ and $f_{L_y|L_x}( \ell_y| L_x=\Delta)$ are continuous functions and unimodal.  
We now have the following lemma:
 \begin{lemma}
 	 If $0 < \ell_y \leq \tfrac{\Delta}{2}$, then $f_{L_y|L_x}(\ell_y | L_x = 0)>f_{L_y|L_x}( \ell_y| L_x=\Delta)$. 
 \end{lemma}
\begin{IEEEproof}
	We first consider the system in \eqref{eq:timingChB} without the absolute value: $\tilde{L}_y = L_x + L_n$. Here, $f_{\tilde{L}_y|L_x}(\tilde{l}_y | l_x) = f_{L_n}(\tilde{l}_y - l_x)$. Since stable distributions are unimodal, we have $f_{\tilde{L}_y|L_x}(\tilde{l}_y | L_x = 0) > f_{\tilde{L}_y|L_x}(\tilde{l}_y | L_x = \Delta), \forall \tilde{l}_y < \frac{\Delta}{2}$. Using the expression for the PDF of system \eqref{eq:timingChB} in \eqref{eq:LyGLx0}--\eqref{eq:LyGLxD} we obtain the desired result. 
	%
	%
	%
	%
\end{IEEEproof}

 \begin{lemma}
 	\label{lemma:existanceThreshold}
 	If $ \ell_y > \tfrac{\Delta}{2}$, then there exists a point $\thr_{\sysB}$ such that for all $\tfrac{\Delta}{2}< \ell_y <\thr_{\sysB}$, $f_{L_y|L_x}(\ell_y | L_x = 0)>f_{L_y|L_x}( \ell_y| L_x=\Delta)$ and for all $\ell_y >\thr_{\sysB}$, $f_{L_y|L_x}(\ell_y | L_x = 0) \leq f_{L_y|L_x}( \ell_y| L_x=\Delta)$. 
 \end{lemma}
 \begin{IEEEproof}
 	Note that for $ \ell_y > \tfrac{\Delta}{2}$, $f_{L_n}(\ell_y) < f_{L_n}(\ell_y-\Delta)$. Moreover, note that  $f_{L_n}(\ell)$ is a smooth function and it is decreasing for $\ell>0$. Then clearly there exists a $\thr_{\sysB} >\tfrac{\Delta}{2}$ such that:
	\vspace{-0.15cm}
	\begin{align}
	\mspace{-5mu} \begin{cases}
	f_{L_n}(\ell_y) \mspace{-3mu} - \mspace{-3mu} f_{L_n}(\ell_y-\Delta) \mspace{-3mu} > \mspace{-3mu}  f_{L_n}(\ell_y+\Delta) \mspace{-3mu} - \mspace{-3mu} f_{L_n}(\ell_y) & \ell_y \mspace{-3mu} < \mspace{-3mu} \thr_{\sysB} \\
	f_{L_n}(\ell_y) \mspace{-3mu} - \mspace{-3mu} f_{L_n}(\ell_y-\Delta) \mspace{-3mu} \leq \mspace{-3mu}  f_{L_n}(\ell_y+\Delta) \mspace{-3mu} - \mspace{-3mu} f_{L_n}(\ell_y) & \ell_y \mspace{-3mu} \geq \mspace{-3mu} \thr_{\sysB},
	\end{cases}		
 	\end{align}
	
	\vspace{-0.1cm}
	\noindent which follows since the slope of $f_{L_n}(\ell)$ for $\ell>0$ decreases, reaches a minimum, and then increases.	Combining both Lemmas, the theorem is proved.
 \end{IEEEproof}
 
\vspace{-0.15cm}
\section{Proof of Theorem \ref{thrm:GpowerStable}} \label{annex:ProofOfGeomPower}

To prove this theorem, we first derive $\mathbb{E}[|N|^s]$. We write this expectation in integral form as:
\vspace{-0.15cm}
\begin{align*}
\mathbb{E}[|N|^s]  &= \int_{-\infty}^{\infty} |n|^s f(n;0,c,\alpha, \beta) dn\\
& \stackrel{(a)}{=}  \int_{0}^{\infty} n^s f(n;0,c,\alpha, \beta) dn \nonumber \\
&\qquad+ \int_{0}^{\infty} n^s f(n;0,c,\alpha, -\beta) dn,
\end{align*}

\noindent where (a) follows since $f(-x;0,c,\alpha, \beta)=f(x;0,c,\alpha, -\beta)$ \cite[Proposition 1.11]{nol15}. Taking the derivative with respect to $s$ we obtain:
\begin{align*}
\frac{d}{ds}\mathbb{E}[|N|^s]  &=  \int_{0}^{\infty} n^s \log n f(n;0,c,\alpha, \beta) dn \nonumber \\
&\qquad+ \int_{0}^{\infty} n^s \log n f(n;0,c,\alpha, -\beta) dn.
\end{align*}

\noindent Further setting $s=0$ results in:
\begin{align*}
\frac{d}{ds}\mathbb{E}[|N|^s]  \bigg\rvert_ {s=0} &= \mathbb{E}[\log(|N|)].
\end{align*}

\noindent We now define $\lambda \triangleq c^\alpha  \sqrt{1+\frac{\beta^2}{\cot^2(\tfrac{\pi \alpha}{2})}}$, 
and let
\begin{align*}
\theta \triangleq 2 \arctan\bigg(\frac{\beta}{\cot(\tfrac{\pi\alpha}{2})} \bigg)/(\pi\alpha).
\end{align*}

\noindent Using \cite[Fact 3, pg. 117]{zol86-book}, and \cite[Theorem 2.6.4]{zol86-book} we have that for $N \sim \StabDist (0,c,\alpha, \beta)$, $\alpha \neq 1$,
\begin{align}
\label{eq:ExpAbsNtoS}
\mathbb{E}[|N|^s]  = \lambda^{s/\alpha} \frac{\cos(\tfrac{\pi}{2} \theta s) \Gamma (1-s/\alpha)}{\cos(\tfrac{\pi}{2} s) \Gamma (1-s)}.
\end{align}
By taking the derivative of \eqref{eq:ExpAbsNtoS} with respect to $s$ and evaluating the result at $s=0$ we obtain:
\begin{align}
 \label{eq:ExpAbsLogN}
 \mathbb{E}[\log(|N|)]  &= \log(c) + \frac{1}{2\alpha}\log\bigg(1+\frac{\beta^2}{\cot^2(\tfrac{\pi \alpha}{2})}\bigg) \nonumber \\
& \mspace{120mu} +(1/\alpha-1)\gamma,
\end{align}

\noindent where $\gamma$ is the Euler's constant \cite[Ch. 5.2]{nist10}. Finally, recalling that $S_0(N) = e^{ \mathbb{E}[\log(|N|)] } $ we conclude the proof. 
 
\section*{Acknowledgment}
The authors would like to thank Professor John P. Nolan at American University for providing valuable correspondence on stable distributions.

\bibliographystyle{IEEEtran}
\bibliography{IEEEabrv,MolCom}

\end{document}